\documentclass[apj]{emulateapj}

\usepackage{multirow}
\usepackage{graphicx}
\usepackage{setspace}
\usepackage{amsmath}
\usepackage{amssymb}

\usepackage{color}
\definecolor{RED}{RGB}{255,0,0}
 \newcommand{\corr}{}

\slugcomment{Accepted for publication in AJ, Nov 27th 2017}

\shorttitle{The Complete transmission spectrum of WASP-39\MakeLowercase{b}}
\shortauthors{Wakeford et al.}

\begin{document}


\title{The Complete transmission spectrum of WASP-39\MakeLowercase{b} with a precise water constraint}


\author{H.R. Wakeford\altaffilmark{1,2}}
\affil{1. Astrophysics Group, University of Exeter, Physics Building, Stocker Road, Devon, EX4 4QL UK}
\affil{2. Space Telescope Science Institute, 3700 San Martin Drive, Baltimore, MD 21218, USA}

\email{stellarplanet@gmail.com}

\author{D.K. Sing\altaffilmark{1}, D. Deming\altaffilmark{3}, N.K. Lewis\altaffilmark{2}, J. Goyal\altaffilmark{1}, T.J. Wilson\altaffilmark{1}, J. Barstow\altaffilmark{4}, T. Kataria\altaffilmark{5}, B. Drummond\altaffilmark{1},  T.M. Evans\altaffilmark{1}, A.L. Carter\altaffilmark{1}, N. Nikolov\altaffilmark{1}, H.A. Knutson \altaffilmark{6}, G.E. Ballester\altaffilmark{7}, A.M. Mandell \altaffilmark{8}}

\affil{3. Department of Astronomy, University of Maryland, College Park, MD 20742, USA}
\affil{4. University College London, London, UK}
\affil{5. NASA Jet Propulsion Laboratory, 4800 Oak Grove Dr, Pasadena, CA 91109, USA}
\affil{6. Division of Geological and Planetary Sciences, California Institute of Technology, Pasadena, CA 91125, USA}
\affil{7. Lunar and Planetary Laboratory, University of Arizona, Tucson, Arizona 85721, USA}
\affil{8. NASA Goddard Space Flight Center, Greenbelt, MD 20771, USA}

\begin{abstract}
WASP-39b is a hot Saturn-mass exoplanet with a predicted clear atmosphere based on observations in the optical and infrared. 
Here we complete the transmission spectrum of the atmosphere with observations in the near-infrared (NIR) over three water absorption features with the Hubble Space Telescope (HST) Wide Field Camera 3 (WFC3) G102 (0.8--1.1\,$\mu$m) and G141 (1.1--1.7\,$\mu$m) spectroscopic grisms. 
We measure the predicted high amplitude H$_2$O feature centered at 1.4\,$\mu$m, and the smaller amplitude features at 0.95 and 1.2\,$\mu$m, with a maximum water absorption amplitude of 2.4 planetary scale heights. 
We incorporate these new NIR measurements into previously published observational measurements to complete the transmission spectrum from 0.3--5\,$\mu$m.
From these observed water features, combined with features in the optical and IR, we retrieve a well constrained temperature T$_{eq}$\,=\,1030$^{+30}_{-20}$\,K, and atmospheric metallicity 151$^{+48}_{-46}\times$\,solar which is relatively high with respect to the currently established mass-metallicity trends. This new measurement in the Saturn-mass range hints at further diversity in the planet formation process relative to our solar system giants.  
\end{abstract}

\keywords{techniques: spectroscopic, planets and satellites: atmospheres, planets and satellites: individual (WASP-39b)}

\section{Introduction}
Exoplanets have greatly advanced our understanding of planetary systems, expanding theory and observations beyond our solar system (e.g. \citealt{seager2010,fortney2010,marley2013,deming2013, stevenson2014b,Sing2016,kataria2016,evans2016,Deming2017}). 
Due to remote observing techniques close-in giant planets dominate atmospheric characterization studies, in part because of their large planet-to-star radius ratios, but mostly due to large extended atmospheres which scatter and transmit a greater number of photons. Observations using transmission spectroscopy have measured atomic (Na and K) and molecular (H$_2$O) absorption in a range of exoplanet atmospheres with ground- and space-based telescopes (e.g. \citealt{charbonneau2002,sing2011a,sing2015a,deming2013,kreidberg2014b,stevenson2014b,nikolov2016,Sing2016,evans2016}). While water has been detected in a number of exoplanet atmospheres, precise and constraining measurements of H$_2$O absorption are still rare (e.g. \citealt{kreidberg2014b,Wakeford2017science}). 

\corr{Observational data analysis is a non-trivial exercise and there are an array of techniques being applied to reduce data (e.g. \citealt{berta2012,gibson2012,deming2013,wakeford2016}).} In a comparative observational study of ten hot Jupiter exoplanets using the Hubble and Spitzer Space Telescopes \corr{with consistent data analysis techniques}, \citet{Sing2016} present a startling diversity in atmospheric transmission spectral features, where clouds play a significant role muting and obscuring atomic and molecular absorption features. \citet{Sing2016} defined a transmission spectral index as a means to distinguish between cloudy and clear atmospheres using the amplitude of water absorption observed with Hubble Space Telescope (HST) Wide Field Camera 3 (WFC3) versus the altitude difference between optical wavelengths, measured with Space Telescope Imaging Spectrograph (STIS), and IR wavelengths, measured with Spitzer Space Telescope Infrared Array Camera (IRAC). The transmission spectral index effectively displayed trends in the `clarity' of an exoplanet atmosphere with a continuum from clear to cloudy for the ten planets in the study. One of the planets in this study is the highly inflated Saturn-mass planet WASP-39b. 
\begin{figure}
\centering 
  \includegraphics[width=0.48\textwidth]{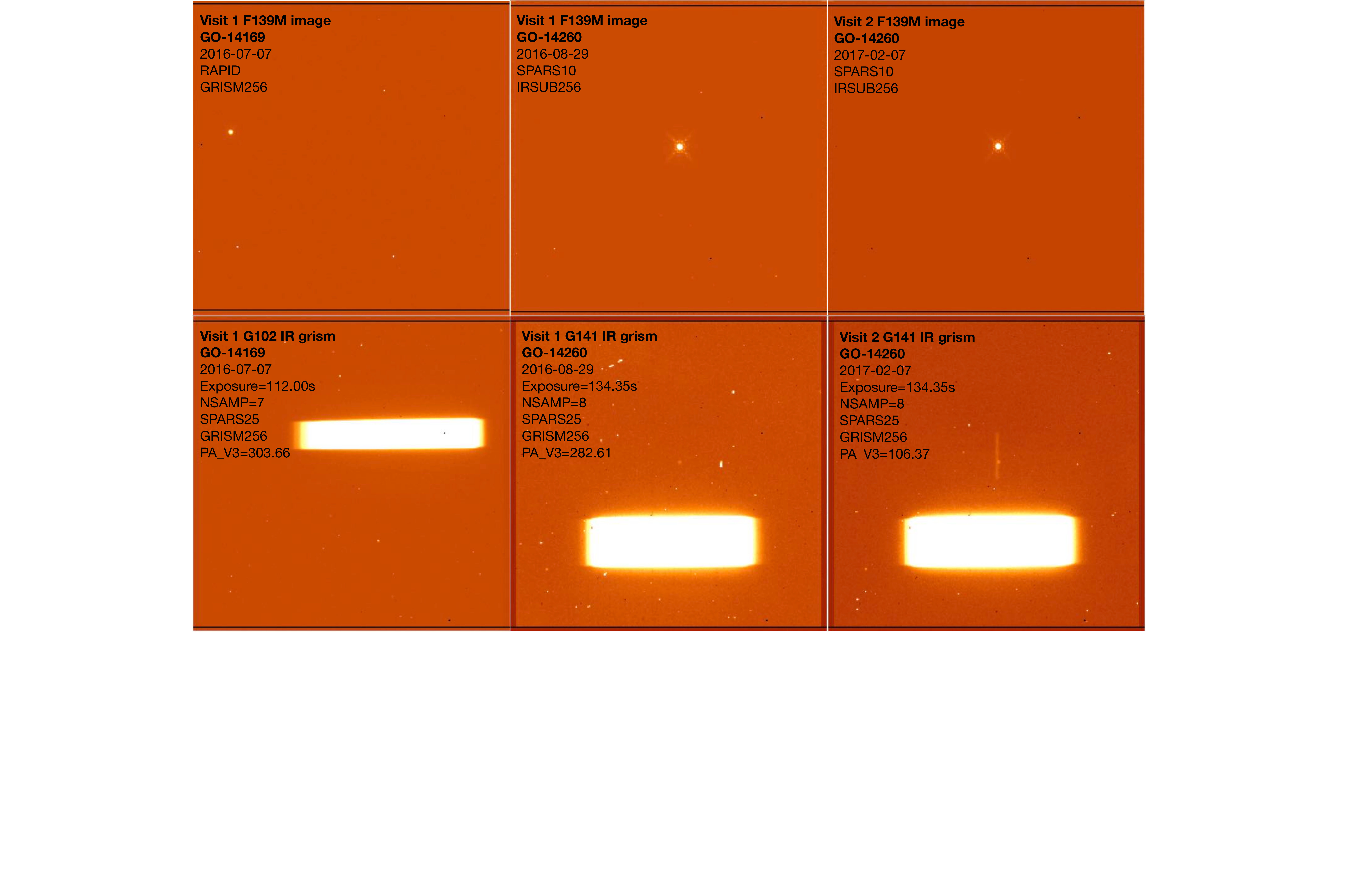}
\caption{\textit{``IMA''} image files of WASP-39 from each visit. The top row shows the direct image taken of the target star with the WFC3 F139M filter, which is used for wavelength calibration of the spectroscopic trace. The bottom row shows the first exposure of each visit: from left to right we show the G102 grism trace, G141 visit 1 trace, and G141 visit 2 trace. Each of the independent observational parameters are listed on the figures. }
\label{fig:W39_ima_images}
\end{figure}

WASP-39b has a radius of 1.27\,R$_J$ and a mass of just 0.28\,M$_J$, and is in orbit around a late G-type star with a period of 4.055 days (\citealt{faedi2011}). Out of current well-characterized exoplanets, only the hot Jupiters WASP-31b and WASP-17b have lower bulk densities (\citealt{anderson2010,anderson2011a,Sing2016}). However, both are more massive and more irradiated than WASP-39b, which has a lower equilibrium temperature of 1116\,K. Analysis of STIS data in the optical and \textit{Spitzer}/IRAC data in the IR shows a spectrum consistent with a predominately clear atmosphere based on both forward model and retrieval studies (\citealt{Sing2016,barstow2017}). The strongest evidence of a clear atmosphere from the current observational data of WASP-39b is the presence of a strong Na feature in the transmission spectrum with pressure broadened line wings for both Na and K absorption lines (\citealt{fischer2016,nikolov2016}). 
These alkali features would be muted or obscured if clouds were present, and would not extend to the H/He continuum. Near-infrared (NIR) features, on the other hand, would be less sensitive to these clouds, given the inherent wavelength dependence of light scattered by small particles (\citealt{bohren2008}).
At the time of the \citet{Sing2016} study, no NIR HST/WFC3 observations existed for WASP-39b, and the atmospheric water absorption could not be measured. Therefore, the `clarity' of the atmosphere (i.e, the transmission spectral index) could not be calculated. In this paper we present three transit observations of WASP-39b using WFC3 from 0.8 to 1.7\,$\mu$m and place precise constraints on the water content in the atmosphere of the planet and calculate a definitive transmission spectral index. 
In section 2 we detail the analysis of the new observations in the NIR. We then present the complete transmission spectrum of WASP-39b from 0.3--5\,$\mu$m in section 3 and introduce the theoretical models used to interpret this dataset. We detail the retrieved atmospheric parameters in section 4 and discuss the implications these have on the nature of WASP-39b with respect to temperature, metallicity, and chemical abundances, as well as planetary formation. 

\begin{figure*}
\centering 
  \includegraphics[width=0.95\textwidth]{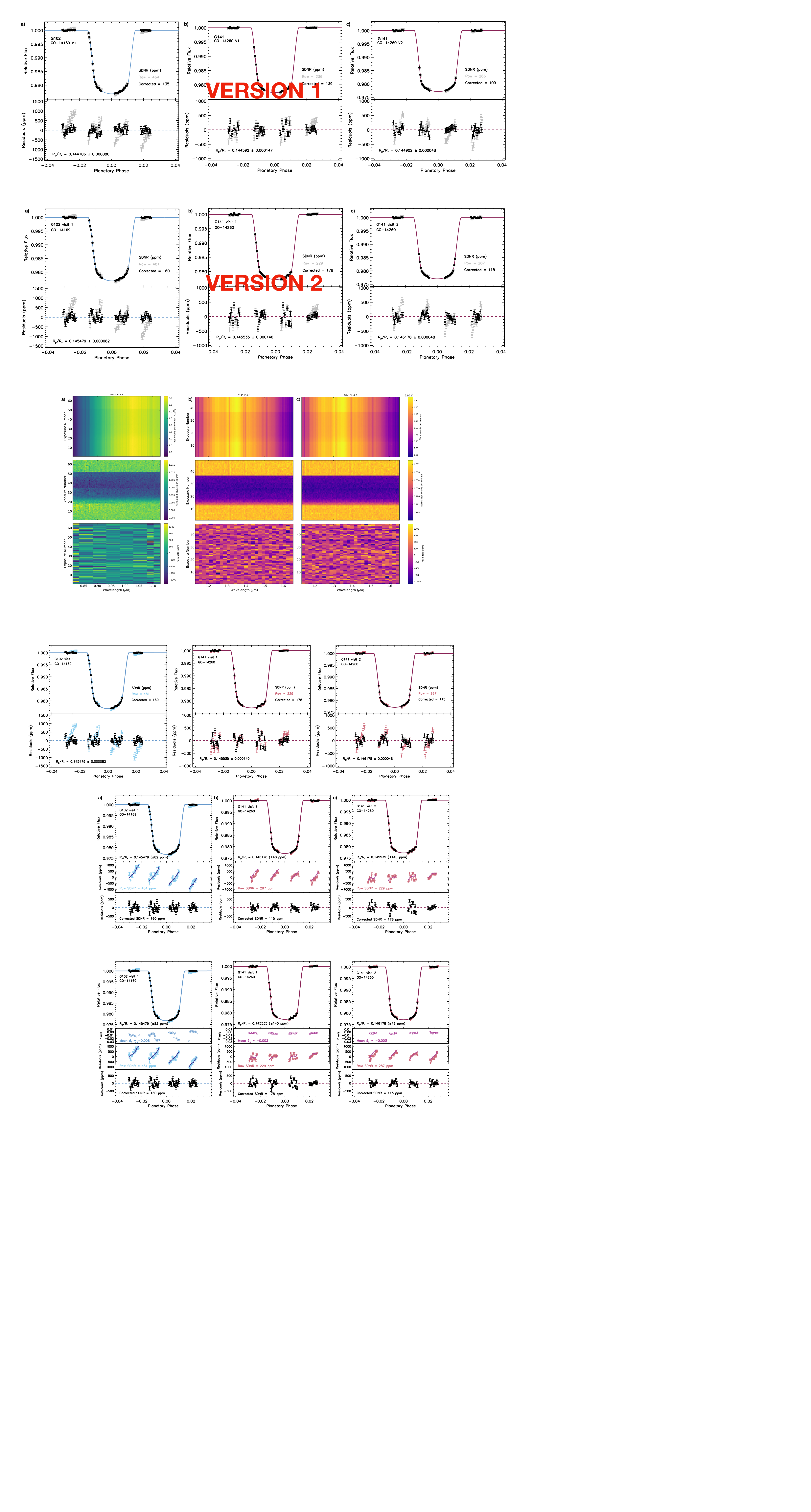}
\caption{White light curves of the WASP-39b transit in each visit. Top: The raw (colored) and corrected (black) lightcurves with the best fit model (solid line). Middle top: $\delta_{\lambda}$ which is the positional shift on the detector between each exposure spectrum. Middle bottom: Raw residuals and uncertainties (colored points), with the best fit systematic model (solid lines). Bottom: Corrected lightcurve residuals (black points). a) G102, b) G141 visit 1, c) G141 visit 2.}
\label{fig:W39_wl}
\end{figure*}

\begin{figure*}
\centering 
  \includegraphics[width=0.95\textwidth]{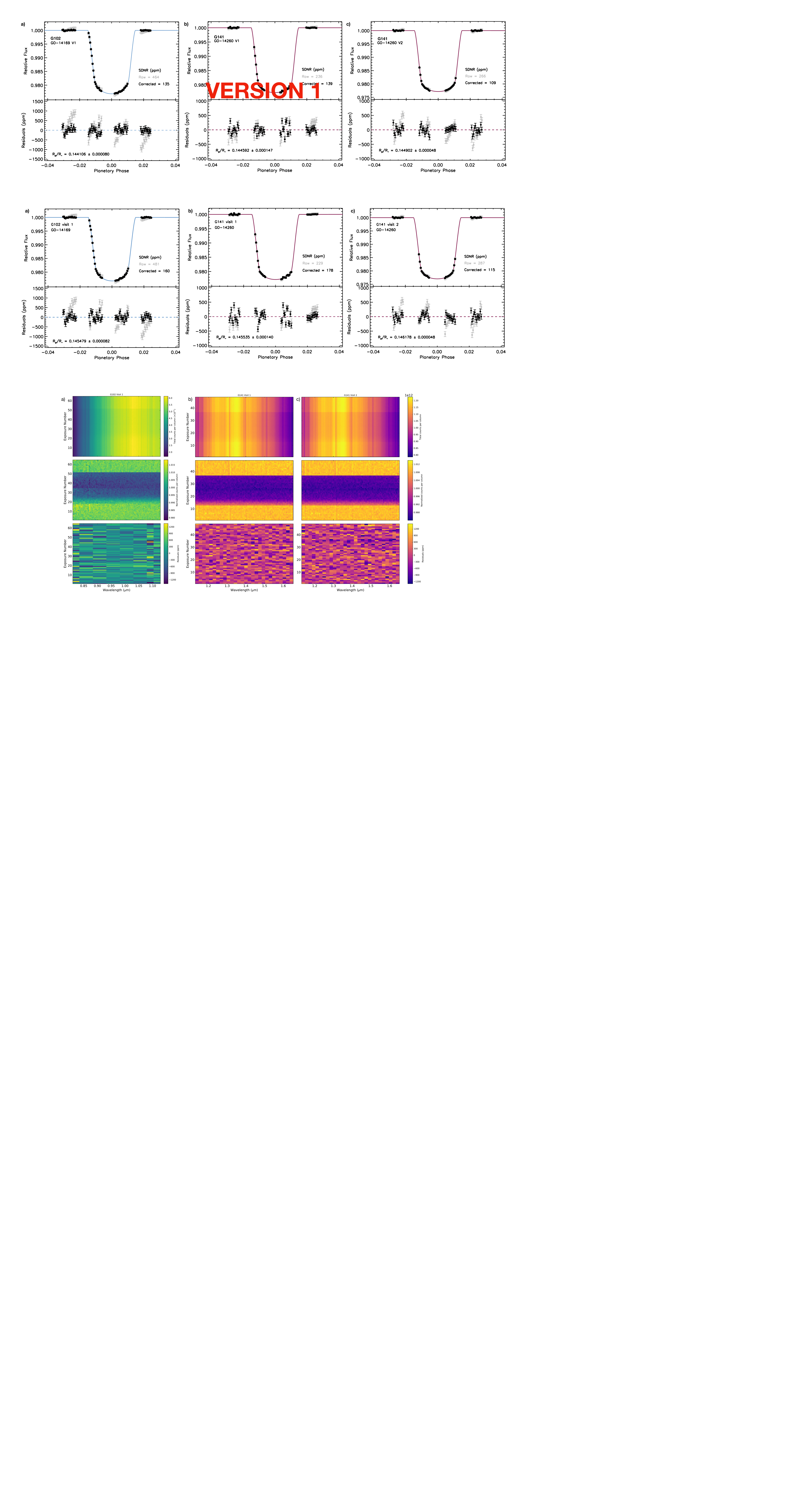}
\caption{Pixel maps for all three observations of WASP-39 with WFC3. Columns a) G102, b) G141 visit 1, and c) G141 visit2. Top row: shows the stellar spectra for each visit following extraction and cosmic ray removal, but prior to systematic corrections, in units of total counts per pixel column. Middle row: shows the stellar spectra normalized by the average counts per wavelength column which enhances the contrast to the in transit exposures. Bottom row: shows the residuals for each spectroscopic light curve after systematic correction.}
\label{fig:W39_jedi}
\end{figure*}

\section{WFC3 Observations and analysis}
We observed WASP-39b during three transit events using HST WFC3 G102 grism as part of GO-14169 (PI. Wakeford) on 7 July 2016, and WFC3 G141 grism as part of GO-14260 (PI. Deming) on 29 August 2016 and 7 February 2017. Observations for both spectroscopic grisms were conducted in forward spatial scan mode. Spatial scanning involves exposing the telescope during a forward slew in the cross-dispersion direction and resetting the telescope position to the top of the scan prior to conducting subsequent exposures. Scans with G102 were conducted at a scan rate of $\sim$0.26 pixels per second with a final scan covering $\sim$28 pixels in the cross-dispersion direction. For G141, we used a scan rate of $\sim$0.30 pixels per second with a final spatial scan covering $\sim$44 pixels in the cross-dispersion direction on the detector (Fig. \ref{fig:W39_ima_images}). 

We use the IMA output files from the CalWF3 pipeline which are calibrated using flat fielding and bias subtraction. We extract the spectrum from each exposure by taking the difference between successive non-destructive reads. A top-hat filter (\citealt{evans2016}) is then applied around the target spectrum and all external pixels are set to zero to aid the removal of cosmic rays (\citealt{nikolov2015}). The image is then reconstructed by adding the individual reads back together. We then extract the stellar spectrum from each exposure with an aperture of $\pm$14 pixels for G102 and $\pm$22 pixels for G141 around a centering profile, which was found to be consistent across the spectrum for each exposure for each observation. 

We monitored each transit with HST over the course of five orbits with observations occurring before, during, and after transit. We discard the first orbit of each visit as it contains vastly different systematics to the subsequent orbits (e.g. \citealt{deming2013,kreidberg2015,wakeford2016,Sing2016,Wakeford2017apjl}). This is due to thermal settling required after target acquisition which results in a different thermal pattern for the telescope not replicated in subsequent orbits. 

\subsection{White lightcurve}
We first analyse the band-integrated light curves of WASP-39, to obtain the broad-band planet-to-star radius ratio (R$_{p}$/R$_{*}$), by summing the flux between 0.8 and 1.13$\mu$m for G102, and between 1.12 and 1.66$\mu$m for G141. The uncertainties on each data point were initially set to pipeline values dominated by photon and readout noise. We account for stellar limb darkening using a 4-parameter limb-darkening law (\citealt{claret2000,sing2010}). Each light curve is fit using the IDL routine MPFIT which uses a Lavenberg-Markwardt (L-M) least-squares algorithm (\citealt{markwardt2009}), which has been shown to produce posterior distributions that are well described by a multi-variable Gaussian distribution (e.g. \citealt{Sing2016,wakeford2016,Wakeford2017science}). 
After an initial fit, the uncertainties on each time series exposure are rescaled based on the standard deviation of the residuals, taking into account any underestimated uncertainties calculated by the reduction pipeline in the data points. The final uncertainty for each point is then determined from the covariance matrix from the L-M algorithm. We fix the system parameters to the previously established values: inclination\,=\,87.36\,$^{\circ}$, $a/R_*$=11.043, period\,=\,4.055259\,days (\citealt{nikolov2016}). From the band-integrated lightcurve we fit for the center of transit time and fix it for subsequent spectroscopic lightcurves. 

\begin{figure}
\centering 
  \includegraphics[width=0.48\textwidth]{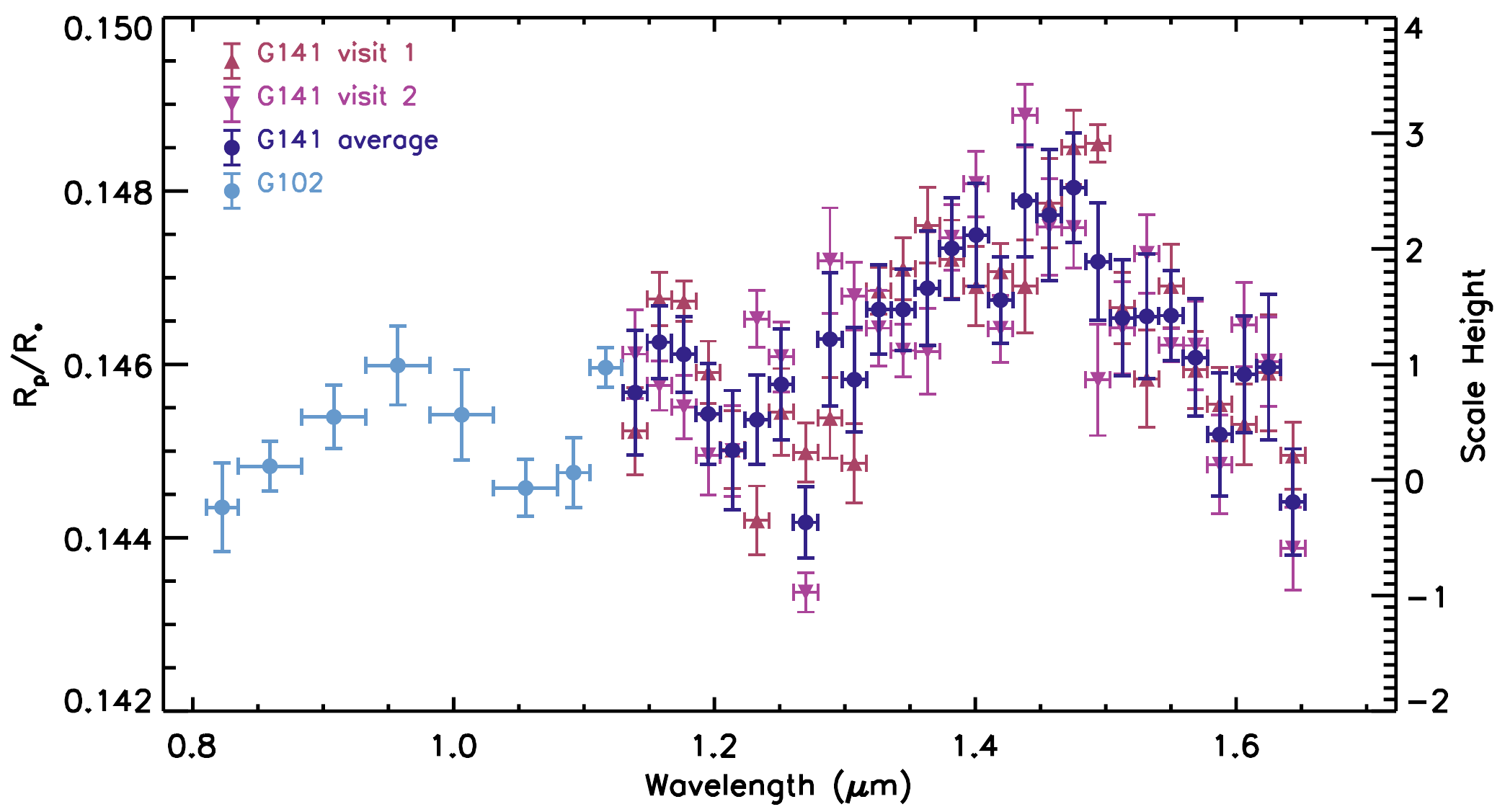}
\caption{WFC3 transmission spectrum measured over three visits with G102 and G141 grisms. The measured transmission spectrum from the G102 visit from 0.8--1.12\,$\mu$m is shown in light blue circles. We analyzed the two visits with G141 separately and show each transmission spectrum (visit 1 triangles, visit 2 downward triangles), and the transmission spectrum computed from the average of the separate spectra (dark red circles). The G141 spectrum is computed from 1.12--1.66\,$\mu$m with a constant bin width of $\Delta\lambda$\,=\,0.0186\,$\mu$m. }
\label{fig:W39_G141_transmission}
\end{figure}

\begin{figure}
\centering 
  \includegraphics[width=0.48\textwidth]{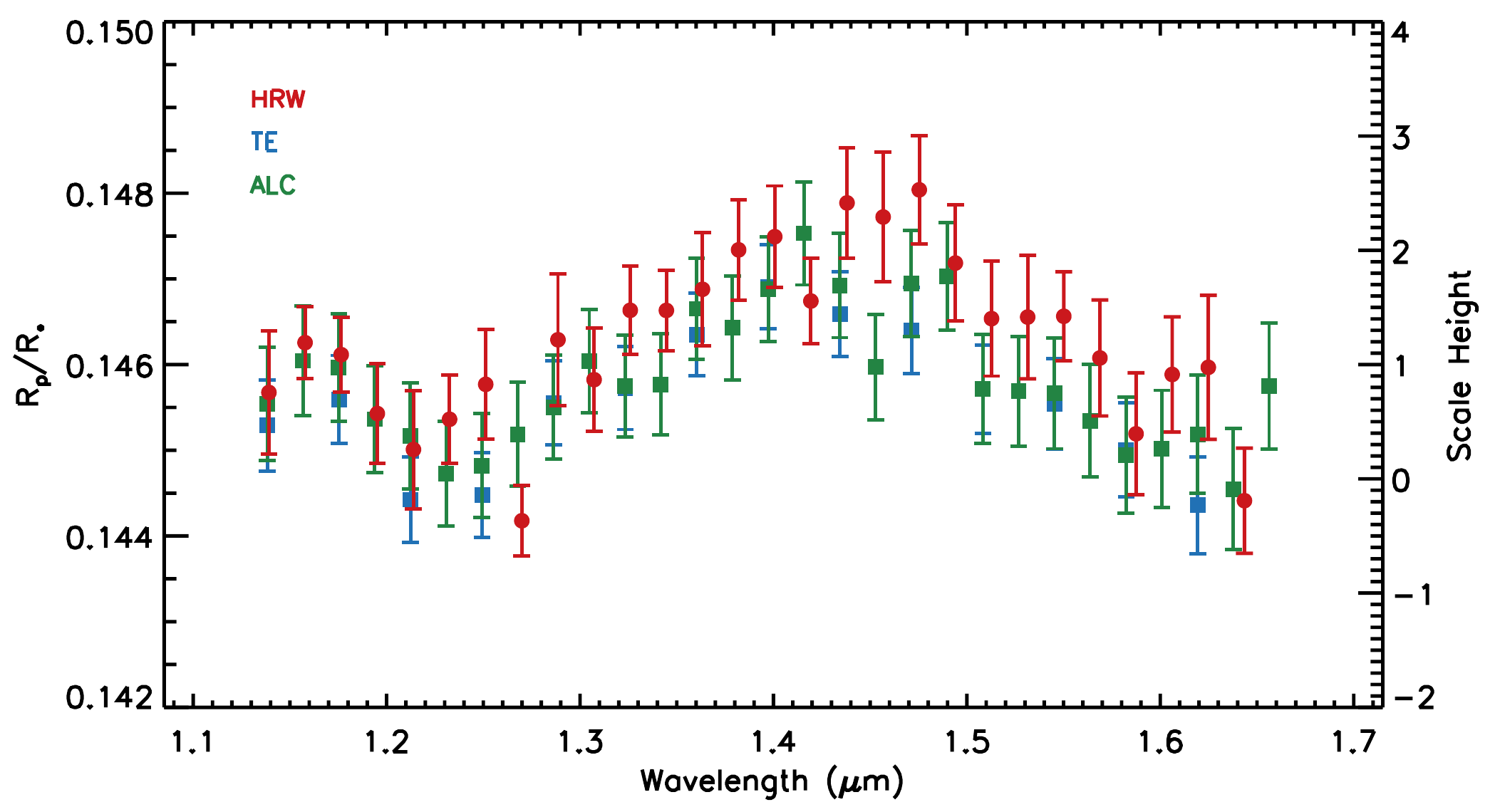}
\caption{Comparison of the average transmission spectrum computed for the G141 visits based on three different analyses. }
\label{fig:W39_G141_comparison}
\end{figure}

We use marginalization across a series of systematic models to account for observatory- and instrument-based systematics (\citealt{gibson2014}). As detailed in \citet{wakeford2016}, we use a grid \corr{of polynomial models,} which approximates stochastic models\corr{, to account for established systematic trends}. Each systematic models accounts for up to three detrending parameters; $\theta$, $\phi$, and $\delta_{\lambda}$, where $\theta$ accounts for a linear increase/decrease of flux in time across the whole observation, $\phi$ is the HST orbital phase trends based on the thermal `breathing' of the telescope over the course of each orbit around the Earth, and $\delta_{\lambda}$ is the shift in wavelength position of the spectrum caused by telescope pointing. We use combinations with and without $\theta$, and with or without $\phi$ and $\delta_{\lambda}$, up to the 4th order. This results in a grid of 50 systematic models which we test against our data, where the most complex has the following function: 
\begin{equation}
S(t,\lambda) = T_1\,\theta \times \sum_{i=1}^{n}p_i\phi^i \times \sum_{j=1}^{n}l_j \delta_{\lambda}^j \mathrm{.}
\end{equation}
Here, T$_1$, p$_i$, and l$_j$ are fit variables or fixed to zero if unfit in a specific model (see \citealt{wakeford2016} for a table all parameter combinations). \corr{It is important to note that marginalization relies on the fact that at least one of the models being marginalized over is a good representation of the systematics in the data. Marginalizing over a series of models then allows for greater flexibility in satisfying this condition, over the use of a single systematic model.}
We use the maximum likelihood estimation (MLE) based on the Akaike Information Criterion (AIC) to approximate the evidence-based weight for each systematic model (\citealt{burnham2004}). \corr{While the AIC is an approximation to the evidence, it allows for more flexible models to be folded into the likelihood, as compared to the Bayesian Information Criterion (BIC), which typically leads to more conservative error estimates on the marginalized values (\citealt{gibson2014,wakeford2016}).} We marginalize across all systematic models to compute the desired light curve parameters. Using marginalization across a large grid of models allows us to account for all tested combinations of systematics and obtain robust center of transit times from the band-integrated light curve, and transit depths for each spectroscopic light curve. We measure a center of transit time of 2457577.425837$\pm$0.000044\,(JD$_{UTC}$) for the G102 visit, 2457630.149461$\pm$0.000044\,(JD$_{UTC}$) for visit 1 with G141, and 2457792.356338$\pm$0.000041\,(JD$_{UTC}$) for visit 2 with G141. We show the white light curves for each visit in Fig.\ref{fig:W39_wl} along with the highest weighted model from our analysis and the marginalized broadband transit depth in terms of $R_p/R_*$. To demonstrate the noise properties of the data, we show the pixel maps of the stellar spectra following spectral extraction, cosmic ray corrections, and wavelength alignment (Fig.\ref{fig:W39_jedi}) in both total counts (top) and in normalized counts (middle). The bottom panel of Fig.\ref{fig:W39_jedi} shows the residuals from each spectroscopic light curve, the analysis of which is outlined in the following section. These plots demonstrate that there are no obvious bad pixels in these data, and no wavelength-dependent trends in the spectroscopic lightcurves.   

\subsection{Spectroscopic light curves}
We divide the WFC3 wavelength range of each grism into a series of bins and measure the R$_p$/R$_*$ from each spectroscopic light curve (Table \ref{table:observation_parameters}) following the same procedure as detailed for the band-integrated light curve. 
For each visit we test a range of bin widths and wavelength ranges. For G102, we test bin widths from $\Delta\lambda$\,=\,0.0146--0.0490\,$\mu$m over wavelength ranges of 0.8--1.12\,$\mu$m and 0.81--1.13\,$\mu$m. For G141, we test bins of $\Delta\lambda$\,=\,0.0186, 0.0128, 0.0373, and 0.0467\,$\mu$m, over two wavelength ranges (1.12--1.65\,$\mu$m 1.13--1.66\,$\mu$m). For each bin and wavelength range we find that the shape of the transmission spectrum is robust and consistent.

For the presented transmission spectrum we divide the G102 spectroscopic range into 8 bins between 0.81 and 1.13\,$\mu$m with variable bin sizes of $\Delta\lambda$\,=\,0.0244 or 0.0490\,$\mu$m (Table \ref{table:observation_parameters}). 
For each spectroscopic lightcurve we test all 50 models and marginalize over the transit parameters to compute the R$_p$/R$_*$. We find that there are few wavelength-dependent systematics as evidenced by the similarity between the highest weighted systematic model for each spectroscopic lightcurve fit.

For the G141 grism observations we extract and analyse each visit separately. As with the band-integrated analysis, we marginalize over all systematic models to compute the transit parameters. For both G141 visits, we use 28 equal bin sizes of $\Delta\lambda$\,=\,0.0186$\mu$m between 1.13 and 1.66$\mu$m (Table \ref{table:observation_parameters}). We found both transmission spectra using the G141 grism to be consistent in absolute depth (Fig. \ref{fig:W39_G141_transmission}). We then combine them into a single spectrum by taking the weighted mean of each spectroscopic measurement (Table \ref{table:observation_parameters}). 

To determine if the analysis method used is justified we compute the transmission spectrum from each visit using a different analysis pipeline (\citealt{evans2016}) over multiple bin sizes (T. Evans (TE): $\Delta\lambda$=0.0500\,$\mu$m, A.L. Carter (ALC): $\Delta\lambda$=0.0186\,$\mu$m) and compute the weighted mean spectrum. We find that the different methods result in the same transmission spectrum in shape and absolute depth within the 1$\sigma$ uncertainties (Fig. \ref{fig:W39_G141_comparison}).
The additional analyses performed used similar techniques to extract the stellar spectra, however, differ in their lightcurve analysis. TE and ALC used Gaussian process (GP) analysis with a Matern v=3/2 kernel (see Evans et al. 2017) on the white light curve to obtain the common-mode systematics. For each spectroscopic light curve, TE used the white ligthcurve systematics and divided each spectroscopic lightcurve by the residuals with the addition of simple linear corrections in time to fit to each wavelength bin (e.g. \citealt{deming2013,evans2017}). Following GP analysis of the white lightcurve, ALC used the stretch and shift method to remove common-mode systematics from each spectroscopic lightcurve (see \citealt{deming2013,wakeford2016}). Both additional methods were computed at different bin sizes and wavelength positions to further test the robust nature of the computed transmission spectrum (Fig. \ref{fig:W39_G141_comparison}). 

\begin{table}
\centering
\caption[\quad HST WFC3 transmission spectrum]{Marginalized transmission spectrum of WASP-39\MakeLowercase{b} measured with HST WFC3 G102 and G141 grism. $\lambda$ marks the center of the bin with $\Delta\lambda$ representing the total width of the bin.}
\begin{tabular}{cccc}
\hline
\hline
$\lambda$ & $\Delta\lambda$ & R$_p$/R$_*$ & Uncertainty \\
$\mu$m & $\mu$m & ~ & (ppm) \\
\hline
\multicolumn{4}{c}{-- G102 --}\\
  0.8225 & 0.0244 & 0.14435 & 310 \\  
  0.8592 & 0.0490 & 0.14482 & 190 \\
  0.9082 & 0.0490 & 0.14539 & 140  \\
  0.9572 & 0.0490 & 0.14598 & 150  \\
  1.0062 & 0.0490 & 0.14541 & 130  \\
  1.0552 & 0.0490 & 0.14457 & 150  \\
  1.0920 & 0.0244 & 0.14475 & 220  \\
  1.1165 & 0.0244 & 0.14596 & 240  \\
 \multicolumn{4}{c}{-- G141 --}\\
  1.1391 & 0.0186 & 0.14567 & 710  \\
  1.1578 & 0.0186 & 0.14625 & 410  \\
  1.1765 & 0.0186 & 0.14611 & 430  \\
  1.1951 & 0.0186 & 0.14542 & 580  \\
  1.2138 & 0.0186 & 0.14500 & 680  \\
  1.2325 & 0.0186 & 0.14536 & 510  \\
  1.2512 & 0.0186 & 0.14576 & 640  \\
  1.2699 & 0.0186 & 0.14417 & 400  \\
  1.2885 & 0.0186 & 0.14628 & 760  \\
  1.3072 & 0.0186 & 0.14582 & 602  \\
  1.3259 & 0.0186 & 0.14663 & 515  \\
  1.3446 & 0.0186 & 0.14663 & 469  \\
  1.3633 & 0.0186 & 0.14687 & 660  \\
  1.3819 & 0.0186 & 0.14733 & 580  \\
  1.4006 & 0.0186 & 0.14749 & 592  \\
  1.4193 & 0.0186 & 0.14674 & 500  \\
  1.4380 & 0.0186 & 0.14788 & 650  \\
  1.4567 & 0.0186 & 0.14772 & 758  \\
  1.4753 & 0.0186 & 0.14803 & 633  \\
  1.4940 & 0.0186 & 0.14718 & 677  \\
  1.5127 & 0.0186 & 0.14653 & 670  \\
  1.5314 & 0.0186 & 0.14655 & 721  \\
  1.5501 & 0.0186 & 0.14656 & 518  \\
  1.5687 & 0.0186 & 0.14607 & 677  \\
  1.5874 & 0.0186 & 0.14519 & 711  \\
  1.6061 & 0.0186 & 0.14588 & 671  \\
  1.6248 & 0.0186 & 0.14596 & 840  \\
  1.6435 & 0.0186 & 0.14441 & 614  \\
\hline
\end{tabular}
\label{table:observation_parameters}
\end{table}

We use these new transmission spectral measurements from G102 and G141 between 0.8--1.66\,$\mu$m (Table \ref{table:observation_parameters}) to complete the transmission spectrum of WASP-39b from the optical to the IR. 

\begin{figure*}
\centering 
  \includegraphics[width=0.98\textwidth]{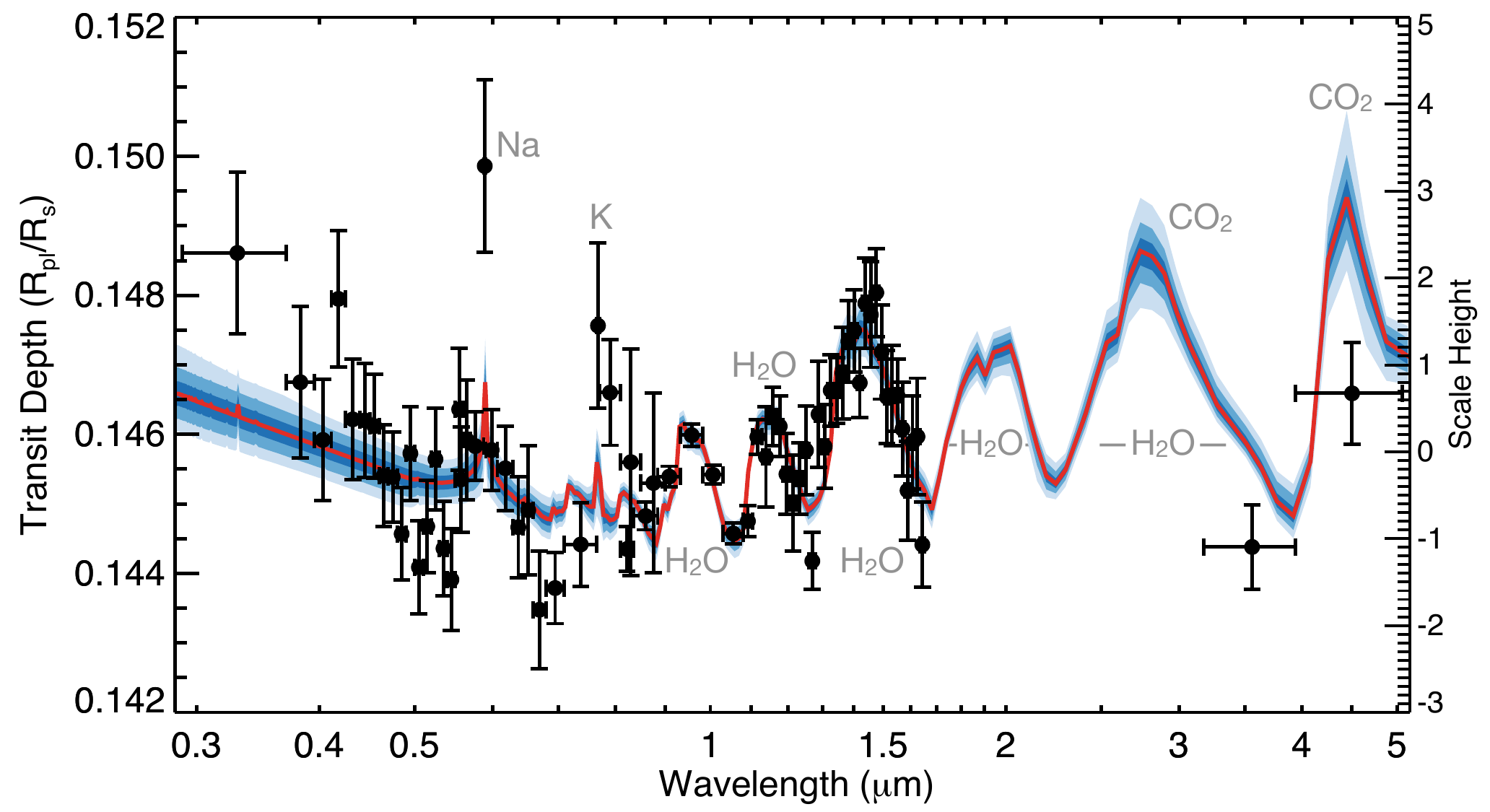}
\caption{The complete transmission spectrum of WASP-39b (black points). This transmission spectrum incorporates data from HST STIS and WFC3, \textit{Spitzer} IRAC, and VLT FORS2 completing the spectrum from 0.3--5.0\,$\mu$m with currently available instruments. Using the ATMO retrieval code, which implements an isothermal profile and equilibrium chemistry, we determine the best fit atmospheric model (red) and show the 1, 2, and 3$\sigma$ confidence regions (dark to light blue) based on the retrieved parameters. }
\label{fig:W39_ARCeq_spec}
\end{figure*}

\vspace{20pt}
\section{WASP-39\MakeLowercase{b's} Complete Transmission Spectrum}
We detect distinct H$_2$O absorption in three bands centered at 0.9, 1.15, and 1.4\,$\mu$m in the new HST WFC3 transmission spectral data (Fig. \ref{fig:W39_ARCeq_spec}). We combine these directly with the previously published HST STIS and \textit{Spitzer} IRAC data (\citealt{Sing2016}), and VLT FORS2 data (\citealt{nikolov2016}), without the need of an offset in absolute depth. As the VLT FORS2 measurements match in wavelength position and depth to the HST STIS data between 0.41--0.81\,$\mu$m (\citealt{nikolov2016}) we take the weighted mean between these measurements, and again do not apply an offset in the absolute depth measured. An offset between datasets can be caused by stellar or planetary variability. However, for an inactive star like WASP-39A (R$_{HK}$\,=\,-4.994), it is more likely caused by differences in the reduction and analysis method used. In our method we apply a consistent analysis across all datasets using marginalization across systematic model grids. The STIS and FORS2 analyses include a common-mode correction; however, both match exactly in depth and wavelength position with overlapping regions with the new WFC3 data, suggesting that any common mode corrections have no significant impact on the absolute transit depth measured. Additionally, each analysis applied the same system parameters as detailed in section 2.1.

\begin{figure}
\centering 
  \includegraphics[width=0.45\textwidth]{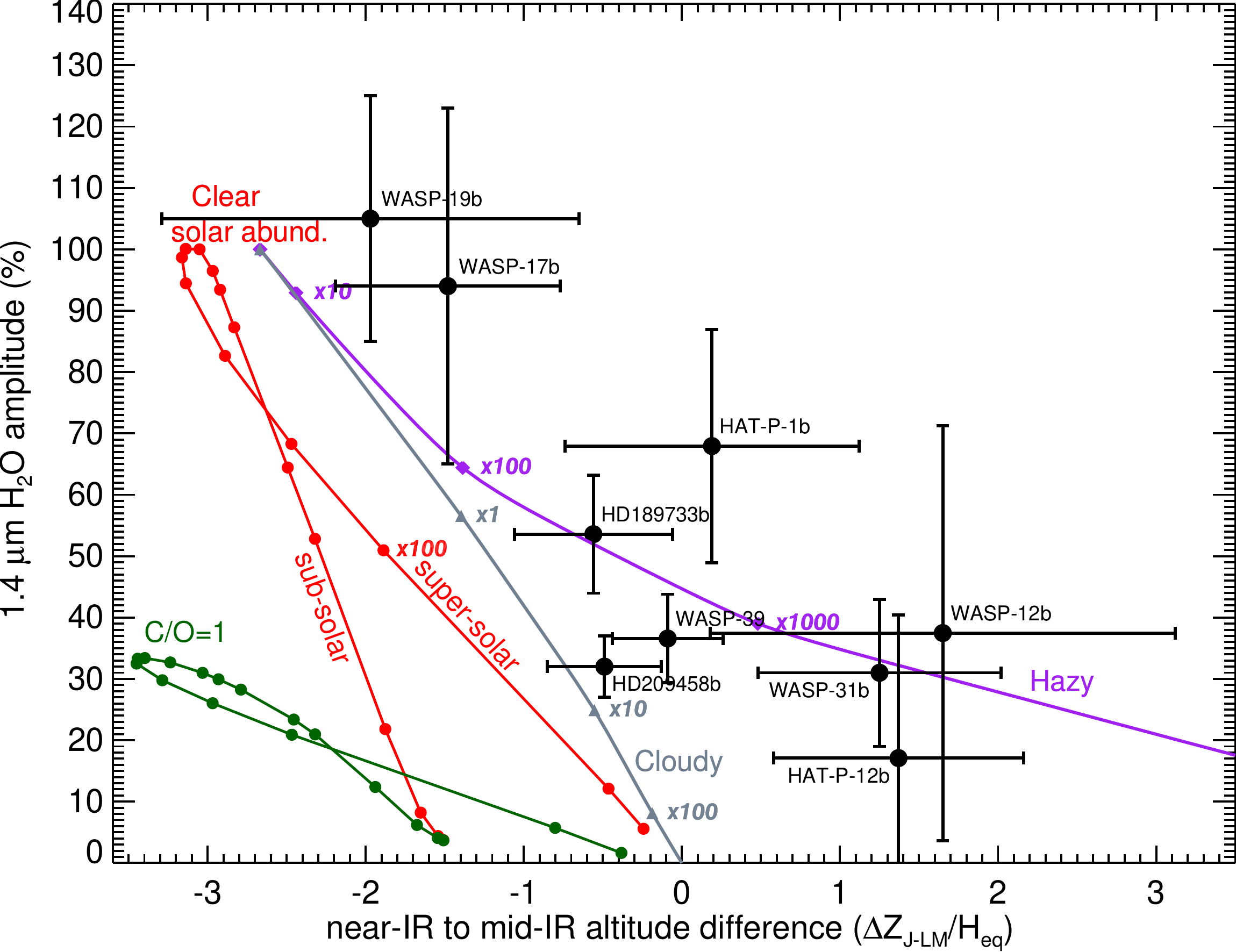}
\caption{Transmission spectral index diagram of $\Delta$Z$_{J-LM}$ versus H$_2$O amplitude as defined in \citet{Sing2016}. Black points show the altitude difference between the  NIR and IR spectral features ($\Delta$Z$_{J-LM}$) versus the H$_2$O amplitude measured at 1.4\,$\mu$m, each with 1$\sigma$ errorbars. Purple and grey lines show the model trends for haze and clouds respectively. Red shows the trend for sub- and super-solar metallicities. Green shows the model trend for different C/O.  }
\label{fig:W39_spec_index}
\end{figure}

We use the complete transmission spectrum of WASP-39b to interpret this inflated Saturn's atmosphere through forward models, retrievals, and 3D GCM simulations. In the following subsections we outline each of these theoretical models and discuss the implications of the results in \S4.

\subsection{Transmission spectral index}
We can now compare WASP-39b to the \citet{Sing2016} sample by computing the transmission spectral index from the H$_2$O amplitude and the $\Delta$Z$_{J-LM}$/H$_{eq}$ altitude difference (see methods section of \citealt{Sing2016}). \corr{The H$_2$O amplitude is calculated by assuming a clear solar atmospheric model, with no cloud or haze opacities, and scaling it in amplitude to fit the observed feature. The $\Delta$Z$_{J-LM}$/H$_{eq}$ altitude difference compares the relative strength of the continuum in the near-IR (1.22--1.33\,$\mu$m) to the mid-IR absorption calculated from the average of the two \textit{Spitzer} points.} We measure an H$_2$O amplitude of 37$\pm$7\% with $\Delta$Z$_{J-LM}$/H$_{eq}$\,=\,-0.09$\pm$0.35. A near zero altitude difference and relatively low water amplitude places WASP-39b further away from the clear solar abundance model, although this does not change its position in the clear to cloudy continuum displayed in \citet{Sing2016} as compared to the other nine exoplanets in the study.
While this index shows that the water amplitude is less than expected for a completely clear solar abundance atmosphere, 
each of these model tracks varies only a single parameter, with all others held fixed; in reality, each of these parameters will vary simultaneously. 
Additionally, previous observations and theoretical models of extrasolar gas giant planets show that these planets may have metallicities greater than 1$\times$ solar (e.g. \citealt{kreidberg2014b,thorngren2016,Wakeford2017science}). We comment further on this planet's metallicity in \S\ref{metallicity}.

\subsection{Goyal Forward Model Grid}
We use the newly developed open source grid of forward model transmission spectra produced using the 1D radiative-convective equilibrium model ATMO (\citealt{Amundsen2014,tremblin2015b,tremblin2016,drummond2016}) outlined in \cite{goyal2017}, to compare to our measured transmission spectrum. Each model assumes isothermal pressure-temperature (P-T) profiles and equilibrium chemistry with rainout condensation. It includes multi-gas Rayleigh scattering and high temperature opacities due to H$_2$O, CO$_2$, CO, CH$_4$, NH$_3$, Na, K, Li, Rb, Cs, TiO, VO, FeH, PH$_3$, H$_2$S, HCN, C$_2$H$_2$, SO$_2$, as well as H$_2$-H$_2$, H$_2$-He collision-induced absorption (CIA). The grid consists of 6,272 model transmission spectra specifically for WASP-39b (i.e. with gravity, R$_p$, etc. of WASP-39b), which explores a combination of eight temperatures (516\,K, 666\,K, 741\,K, 816\,K, 966\,K, 1116\,K, 1266\,K, and 1416\,K), seven metallicities (0.005, 0.1, 1, 10, 50, 100, 200$\times$ solar), seven C/O values (0.15, 0.35, 0.56, 0.70, 0.75, 1.0, and 1.5), four ``haze'' parameters (1, 10, 150, and 1100), and four cloud parameters (0, 0.06, 0.2, and 1). In the Goyal grid the ``haze'' parameter defines an enhanced Rayleigh-like scattering profile, which increases the hydrogen cross section with a wavelength dependent profile. The ``cloud'' parameter defines a grey uniform scattering profile across all wavelengths between 0 and 100\% cloud opacity (see \citealt{goyal2017} for more details). 

We fit each model to the transmission spectrum by only allowing them to move in absolute altitude; we therefore have one free parameter for each model fit. We use the L-M routine MPFIT to determine the best fit altitude and calculate the $\chi^2$. We transform from $\chi^2$ to probability likelihood via the expression 
\begin{equation}
p\left(\textbf{x}\right) = A \exp\left(-\frac{1}{2} \chi^2(\textbf{x})\right),
\label{eq:probfromchi2}
\end{equation}
where $\textbf{x} \equiv \{T,\ \mathrm{[M/H]},\ \mathrm{C/O},\ H,\ C\}$ are the temperature, metallicity, carbon-to-oxygen ratio, haze, and cloud parameters respectively and $A$ is a normalization constant. This constant is set by integrating the likelihood over all model parameters, such that
\begin{equation}
\idotsint p\left(\textbf{x}\right)\,\mathrm{d}\textbf{x} = 1.
\label{eq:normprob}
\end{equation}
The Goyal grid is specifically generated for each planet and the range of each of the parameters are reasonable assumptions for a planetary atmosphere. As such we assume a weakly informative prior on each model, which is uniform to the edge of the grid space where it is truncated. 

\begin{figure}
\centering 
  \includegraphics[width=0.45\textwidth]{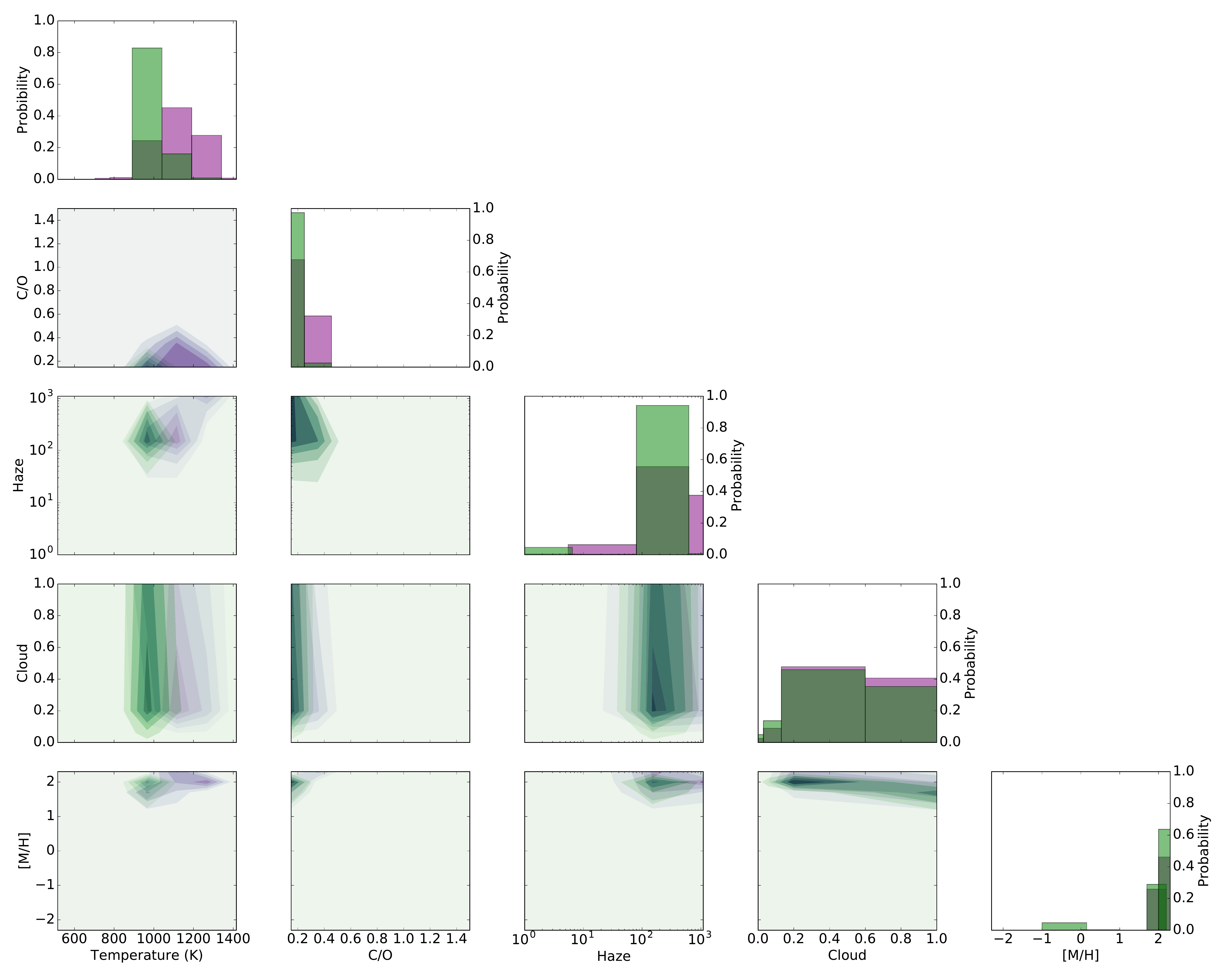}
\caption{Probability distribution of the forward model grid fit to the data for the new WFC3 results only (purple) and the full transmission spectrum including STIS, VLT, and \textit{Spitzer} points (green). These pairs plots show the correlations between the five parameters explored in the model grid with the respective probability histograms for each parameter. }
\label{fig:W39_grid_cp}
\end{figure}

\begin{figure}
\centering 
  \includegraphics[width=0.45\textwidth]{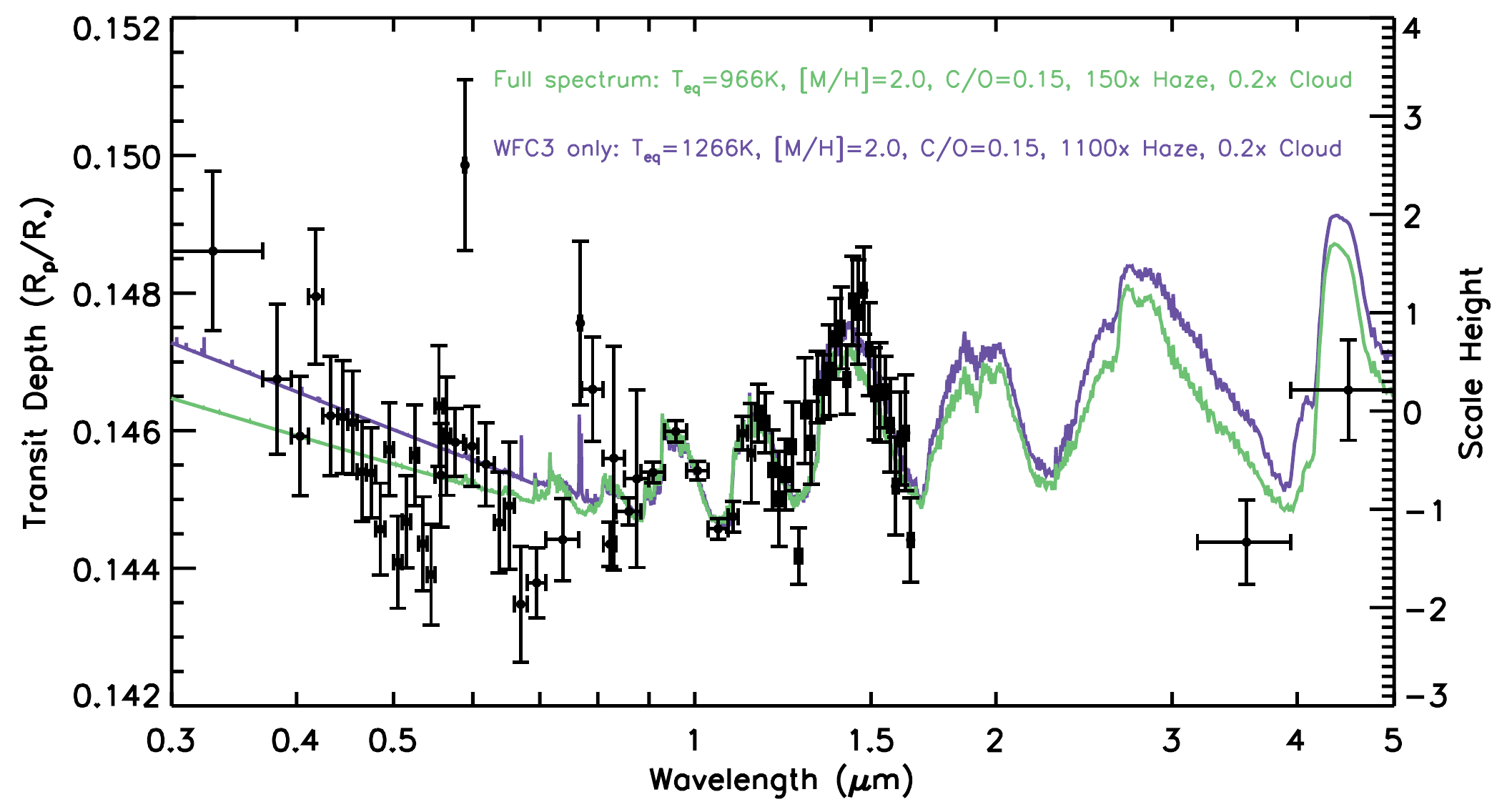}
\caption{Highest probability grid models fit to the the transmission spectrum for, the new WFC3 results only (purple) and the full transmission spectrum (green) including STIS, VLT, and \textit{Spitzer} points.}
\label{fig:W39_grid_spectrum}
\end{figure} 

We fit the model grid to both the full transmission spectrum (0.3--5\,$\mu$m) and a subset containing only the new WFC3 data (0.8--1.7\,$\mu$m). The probability distribution of the model grid is shown in Fig.\,\ref{fig:W39_grid_cp} for both scenarios. We then select the model with the highest probability for each dataset and plot them in Fig.\,\ref{fig:W39_grid_spectrum} against the complete transmission spectrum of WASP-39b. From these plots we see a distinct difference between the best fit temperature and haze parameters with just the H$_2$O features, and with the full optical and NIR transmission spectrum. These differences predominantly impact the optical portion of the spectrum which anchors the haze parameter, but is largely degenerate with the temperature. Additionally, the haze and cloud parameters have a widespread influence on the C/O and metallicity, which is further highlighted by the inclusion of the optical and IR data. We discuss this further in \S\ref{discussion}.  

\begin{table*}
\centering
\caption[\quad ATMO retrieval models]{Results from each retrieval model}
\begin{tabular}{ccccccccc ccc}
\hline
\hline
Data& Model  &k& DOF & $\chi^2$ & BIC & T$_{eq}$ & M/H & C/O & H$_2$O & Na & K \\
 ~ & Chemistry &~ &  ~ & ~ & ~ & (K) & [log$_{10}$] & ~ &[VMR]&[VMR]&[VMR] \\
 ~ & ~ & ~ & ~ & ~ & ~ & ~ & $\times$solar & ~ &$\times$solar&$\times$solar&$\times$solar \\
\hline
\\
WFC3 & Equilib. &5&  31 & 31.9 & 49.8 & 950$^{+130}_{-150}$ & 1.88$^{+0.41}_{-0.65}$ & 0.19$^{+0.29}_{-0.16}$\\
 ~ & ~ & ~ &  ~ & ~ & ~ & ~ & 76$^{+117}_{-59}$ & ~\\
Full & Equilib. &5&  65 & 88.1 & 109.0 & 1030$^{+30}_{-20}$ & 2.18$^{+0.12}_{-0.16}$ & 0.31$^{+0.08}_{-0.05}$\\
 ~   &              ~ &  ~ &   ~ &     ~ &      ~ &                  ~ & 151$^{+48}_{-46}$ & ~ \\
Full & Free-     &10&  60 & 85.8 & 128.3 & 920$^{+70}_{-60}$ &2.07$^{+0.05}_{-0.13}$ &0.44$^{+0.11}_{-0.44}$ & -1.37$^{+0.05}_{-0.13}$ & -3.7$^{+0.6}_{-0.6}$ & -4.9$^{+0.8}_{-1.3}$\\
   ~ & Chem. & ~ &   ~ &     ~ &      ~ &                 ~ & 117$^{+14}_{-30}$ & ~ & 117$^{+14}_{-30}$  ~&120$^{+330}_{-90}$ &105$^{+620}_{-100}$\\
\\
\hline
\multicolumn{7}{l}{[VMR] is log$_{10}$(Volume Mixing Ratio)}
\end{tabular}
\label{table:retrievals}
\end{table*}

\subsection{ATMO Retrieval}\label{arc_retrieval}
We use the ATMO Retrieval Code (ARC) to fully explore the parameter space covered by the transmission spectral measurements. Using ARC we calculate the posterior distribution of the data to the models, and determine the fit confidence intervals marginalizing over parameter space (see \citealt{Wakeford2017science} and \citealt{evans2017} for further details). ARC couples the ATMO model to a L-M least-squares minimizer and a Differential Evolution Chain Monte Carlo (DECMC) analysis (\citealt{eastman2013}).
We ran DECMCs with 12 chains each with 30,000 steps, with convergence typically occurring around 15,000 steps, where convergence is monitored using the Gelman-Rubin statistic.
We considered two atmospheric retrieval models: a chemically consistent retrieval scheme which assumes chemical equilibrium, and a more flexible free-chemistry retrievals. In the equilibrium chemistry retrieval, the chemical abundances are consistent with the pressure-temperature profile. In the free-chemistry retrieval, the abundances are assumed to be constant with pressure and are independently fit. For the opacity sources, we include H$_2$-H$_2$, H$_2$-He CIA, and molecular opacities from H$_2$O, CO, CO$_2$, CH$_4$, NH$_3$, Na, and K. For both models we assume isothermal P-T profiles. Parameterized P-T profiles were explored, although found to be unnecessary to fit the data, as the retrieved P-T profiles were found to be isothermal over the altitudes probed by the data with no significant structure in the P-T profile indicated by the fit parameters. 

\begin{figure}
\centering 
  \includegraphics[width=0.45\textwidth]{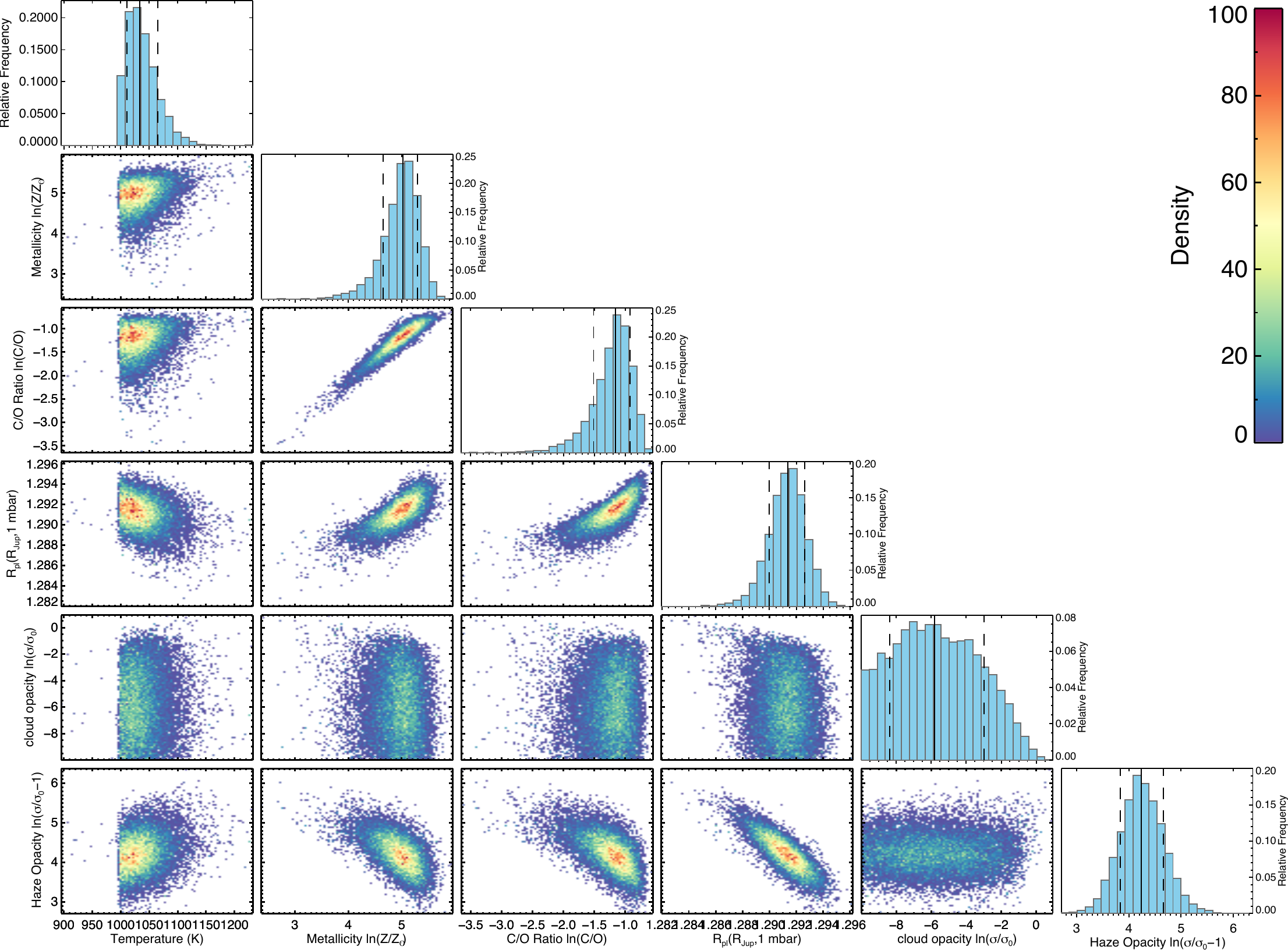}
\caption{Probability density map of the ARC fit to the data for the  the full transmission spectrum, using equilibrium chemistry and an isothermal P-T profile. These pairs plots show the correlations between the six parameters explored in the ARC DECMC with the respective probability density histograms for each parameter. }
\label{fig:W39_ARC_cp}
\end{figure}

In each retrieval we fit for a haze and cloud parameter, each of which are scaling factors applied to the hydrogen cross-section with either a wavelength dependent profile (haze) or uniform profile (cloud) - these two parameters are represented in ARC (Fig. \ref{fig:W39_ARC_cp}) as a natural log ratio for numerical reasons only.  Using the equilibrium retrieval we also run a fit to only the new WFC3 data, as with the Goyal grid, to determine the impact of the optical and IR transmission spectrum in the retrieved parameters. 

In Table~\ref{table:retrievals} we list the main results from the ARC analysis, with each model having the following free parameters, respectively:\\
Equilibrium: planetary radius (R$_p$), T$_{eq}$, Haze, Cloud, [M/H], C/O. \\
Free-chemistry: R$_p$, T$_{eq}$, Haze, Cloud, H$_2$O volume mixing ratio (VMR), CO$_2$ VMR, CO VMR, CH$_4$ VMR, Na VMR, and K VMR.\\
From this retrieval we are able to place constraints on the equilibrium temperature of the observed portion of the planetary atmosphere, as well as the C/O and metallicity by fitting the absorption features. In the equilibrium chemistry case the abundance of elemental carbon and oxygen species is set via the C/O and metallicity. Specifically in ATMO, the carbon abundance is set as a multiple of the solar carbon ($C$) abundance: 
\begin{equation}
A(\mathrm{C}) = A(\mathrm{C}, solar) \times 10^{\mathrm{[M/H]}}, 
\end{equation}
and then the oxygen (O) abundance is set via the carbon abundance and the CO ratio: 
\begin{equation}
A(\mathrm{O}) = A(\mathrm{C}) / CO_{ratio}.
\end{equation}
In the free chemistry model each species abundance is fit for independently, and [M/H] is estimated based on the H$_2$O abundance. The C/O ratio is then estimated from the four fit molecules containing carbon and/or oxygen (CO, CO$_2$, CH$_4$, H$_2$O) and does not take into account any other species. The Na and K line profiles are modeled using an Allard profile (\citealt{allard2007}); alternate profiles were tested but did not improve upon the statistical fit to the data.

\begin{figure*}
\centering 
  \includegraphics[width=0.95\textwidth]{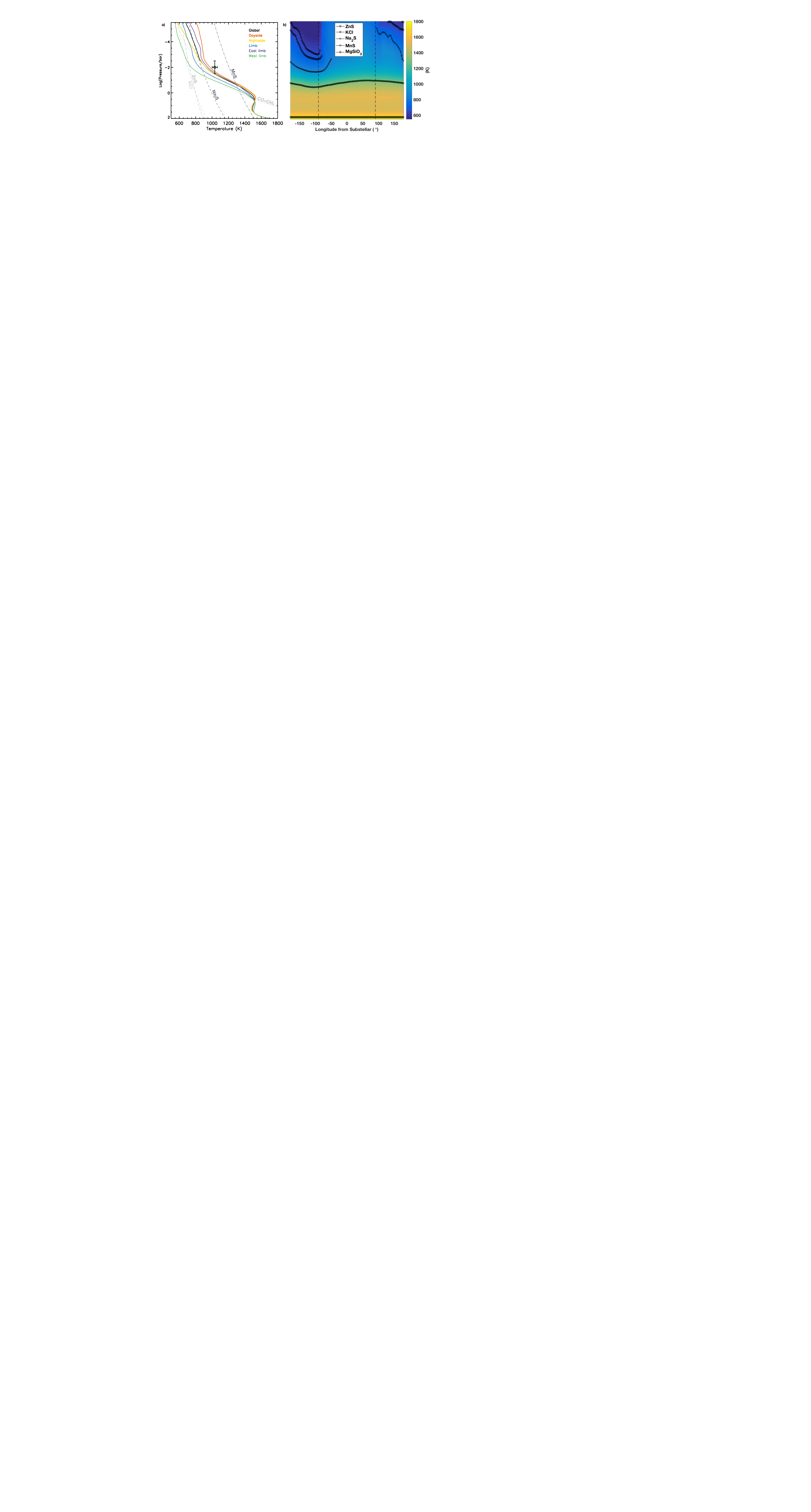}
\caption{a) 1$\times$\,solar 3D P-T profiles of WASP-39b. The global, dayside, nightside, limb, east, and west limb averaged profiles computed from 3D GCM models of WASP-39b (see \citealt{kataria2016}). Also plotted are the condensation curves of the affecting species, as well as the CO/CH$_4$ abundance line. These P-T profiles indicate a difference in temperature expected on opposite limbs of the planet, resulting in the potential for one limb to be observationally clear while the other contains more significant cloud opacities. The point indicates the temperature retrieved for the atmosphere of WASP-39b based on an isothermal equilibrium model (Table \ref{table:retrievals}). 
b) Average temperature of WASP-39b as a function of longitude and pressure. Temperatures are weighted by the cosine of the latitude, equivalent to weighting each grid point by its projection angle toward an observer at the equator. Low-temperature condensates are overplotted in this 2D space.}
\label{fig:TP}
\end{figure*}

We find that the data are described best with an isothermal equilibrium model which has a $\chi^2_\nu$ of 1.32 compared to 1.45 for the free chemistry fit. From this we retrieve T$_{eq}$=1030$^{+30}_{-20}$ and an atmospheric metallicity 151$^{+48}_{-46}\times$\,solar, as defined by the posterior distributions (Fig.\ref{fig:W39_ARC_cp}).

\subsection{3D GCM}\label{gcm}
To model the three-dimensional (3D) temperature structure of WASP-39b we use the SPARC/MITgcm (\citealt{showman2009}). The SPARC/MITgcm couples the MITgcm, a finite-volume code that solves the 3D primitive equations on a staggered Arakawa C grid (\citealt{adcroft2004}) with a two-stream adaptation of a multi-stream radiative transfer code for solar system planets (\citealt{marley1999}). The radiative transfer code employs the correlated-k method with 11 bands optimized for accuracy and computational efficiency. The opacities are calculated assuming local thermodynamic and chemical equilibrium. This code has been used extensively to model the atmospheric circulation of exoplanets (e.g., \citealt{lewis2010,kataria2015,kataria2016,Wakeford2017science,lewis2017}). 

Here we show the P-T profiles from \citet{kataria2016} for WASP-39b averaged over different regions of the atmosphere, demonstrating the impact of 3D circulation on the planetary P-T profiles (Fig. \ref{fig:TP}a). We also plot the retrieved temperature based on isothermal equilibrium models at the pressure probed by these observations. Also shown are the condensation curves of potential cloud species in the atmosphere of WASP-39b, as well as the molecular transition region from CO-dominated carbon chemistry to CH$_4$-dominated carbon chemistry. From the 3D GCM we derive a condensation map showing the average temperature as a function of longitude and pressure indicating where and what condensates might form in the atmosphere (Fig. \ref{fig:TP}b). The GCM results suggest that the two limbs of the planet likely have different cloud properties due to the recirculation of heat around the planet from the dayside hemisphere. The nightside trailing limb (west) is approximately 100\,K colder than the sun trailing limb (east) at pressures of 0.1\,bar, and up to 200\,K colder at pressures less than 1\,mbar. These differences in the limbs not only influence the average temperature observed in transmission, but also in this case will impact the condensate cloud species likely to form in the atmosphere on the colder western limb.
This becomes important when considering the impact differences in the limbs will have on the transmission spectrum which measures the average absorption profile of the limb annulus around the planet (e.g. \citealt{line2016b}).  

We note that this model is for a 1\,$\times$solar composition case and with an increased atmospheric metallicity the temperature is likely to increase as well as the position of the condensation curves (\citealt{wakeford2017mnras}). Additionally, there is evidence that the day-night temperature contrast will also change (likely increase) with metallicity (\citealt{kataria2014,charnay2015}, Drummond et al. in prep), which would have consequences on the east-west limb cloud formation.
 
\section{Discussion}\label{discussion}
We present the complete transmission spectrum of WASP-39b from 0.3--5.0\,$\mu$m combining data from HST STIS and WFC3, VLT FORS2, and \textit{Spitzer} IRAC. Figure \ref{fig:W39_ARCeq_spec} shows the absorption features of WASP-39b's atmosphere which includes broad sodium line wings and three distinct water absorption peaks. There is also tentative evidence of absorption by potassium in the optical and absorption due to carbon-based species in the \textit{Spitzer} 4.5\,$\mu$m channel. 

In our analysis we find that the data are described best with an isothermal equilibrium model with T$_{eq}$\,=\,1030$^{+30}_{-20}$ and [M/H]\,=\,151$^{+48}_{-46}\times$\,solar. At the 1$\sigma$ level WASP-39b has some of the most constrained atmospheric parameters to date. In this section we discuss the implications from the interpretive methods used and the importance of the complete transmission spectrum. 

\subsection{Forward models and retrievals}
To interpret the transmission spectrum of WASP-39b we used a grid of 6,272 forward models that were specifically calculated for WASP-39b (\citealt{goyal2017}). The Goyal grid samples five different atmospheric parameters commonly explored in a retrieval code; temperature, metallicity, C/O, scattering haze, and uniform clouds. 
Based on results from the Goyal grid we find that the data are best described by a high metallicity (100$\times$solar), low C/O (0.15), hazy (150$\times$) atmosphere (Fig. \ref{fig:W39_grid_spectrum}). The large sampling of models in the grid allows for more detailed interpretation compared to taking a handful of non-specific exoplanet models, with the addition that the probability distributions can be explored. 
The grid probability distributions suggest that uniform clouds are not playing a significant role in the observed transmission spectrum, where any cloud parameter applied to the data is equally probable given a distribution of the other four parameters (Fig. \ref{fig:W39_grid_cp}). 
As there are no carbon-bearing species in the transmission spectrum of WASP-39b, the C/O ratio is constrained to the lowest value in the grid. This makes it more difficult to statistically infer precise atmospheric values as the tail end of the distributions are being cut (c.f. \citealt{goyal2017}).

Using the ATMO retrieval code (ARC) we implement an isothermal equilibrium chemistry model and expand upon the parameter space covered by the Goyal grid. While the grid contains sparse sampling of all five parameters, the ARC uses DECMC to throughly explore tens of thousands of models. From both the Goyal grid and ARC we find hard cut-offs in the temperature space; this is likely due to the inclusion of the optical data and the presence of sodium absorption in the observed spectrum, which condenses in the models for lower temperatures (see section\,\ref{opvir}). In the Goyal grid, the temperature cut-off appears at 966\,K, while the finer sampling used in the ARC leads to a more accurate cut-off of $\sim$1000\,K.  Below this temperature, significant amounts of rainout removes Na from the gas phase, which condenses into Na$_2$S and/or NaAlSi$_3$O$_8$. 

Using ARC, we can further explore the distribution of C/O and metallicity. While these parameters are highly correlated, ARC is able to explore the limits of the distribution with C/O\,$<$\,0.15, and [M/H]\,$>$\,2.7. Interestingly, increased sampling on the cloud parameter yields no further interpretation. One advantage of using the Goyal grid over ARC is speed -- the fit to the entire Goyal grid can be run in minutes, while a full retrieval with ARC can take many days. From our analysis, we can see that the Goyal grid is able to constrain planetary parameters within reasonable limits as described by the current data. 

\subsection{Interpreting the chemistry}
Using ARC we run both an equilibrium chemistry model and a free-chemistry model to fit for chemical species and abundances in the measured atmosphere (see \S\ref{arc_retrieval}). 
Both ATMO chemistry models retrieve similar parameters for temperature, metallicity, and C/O (Table \ref{table:retrievals}). The high resolution and precision of data over the 0.9, 1.2, and 1.4\,$\mu$m water absorption features place good constraints on the temperature and metallicity retrieved by each model. These are further reinforced in the free-chemistry fit by the individual abundances retrieved for the Na and K absorption features in the optical, which have similar volume mixing ratios (VMR) to the H$_2$O, and therefore inform the overall atmospheric metallicity. The presence of Na in the transmission spectrum also informs the temperature of the atmosphere, as at low temperatures ($<$1000\,K) Na would likely be depleted by rainout, confining it deeper in the atmosphere. \corr{The thermosphere of an exoplanet is most sensitively probed by the Na line core (e.g. \citealt{huitson2012}). However, the line core of Na is not resolved in the transmission spectrum of WASP-39b, which prohibits a more detailed investigation of the line core and the inclusion of the thermosphere in the models detailed. In the case that the Na line core temperature is underestimated, the model will not be able to match the data higher in the atmosphere (Figs. \ref{fig:W39_ARCeq_spec}, \ref{fig:W39_grid_spectrum}). To further constrain the temperature and structure of WASP-39bs thermosphere, high resolution observations of Na will be required.}

Following the results from the solar-metallicity GCM of WASP-39b (Fig.\,\ref{fig:TP}), we might expect to observe absorption by CH$_4$ in the atmosphere of WASP-39b, as a number of the P-T profiles cross the CO\,=\,CH$_4$ boundary. However, we do not see any evidence for methane absorption in the transmission spectrum, which would be present at the red end of the G141 bandpass ($>$1.55\,$\mu$m) and in the 3.6\,$\mu$m \textit{Spitzer} bandpass. 
Both the equilibrium chemistry and free-chemistry retrievals suggest that the atmosphere of WASP-39b likely has a metallicity greater than 100$\times$solar. The high metallicity will impact not only the P-T profiles by pushing them to higher temperatures but would also push the CO\,=\,CH$_4$ transition to lower temperatures, particularly at high  pressures (\citealt{agundez2014}). This likely places the atmosphere of WASP-39b firmly above the CO\,=\,CH$_4$ boundary, in the CO dominated regime, for all pressures and all locations horizontally.
It is therefore not likely that the CH$_4$ abundance is enhanced relative to equilibrium abundances due to horizontal or vertical quenching, as has been postulated for many hot Jupiter atmospheres (e.g. \citealt{cooper2006,agundez2014}).

\begin{figure}
\centering 
	\includegraphics[width=0.45\textwidth]{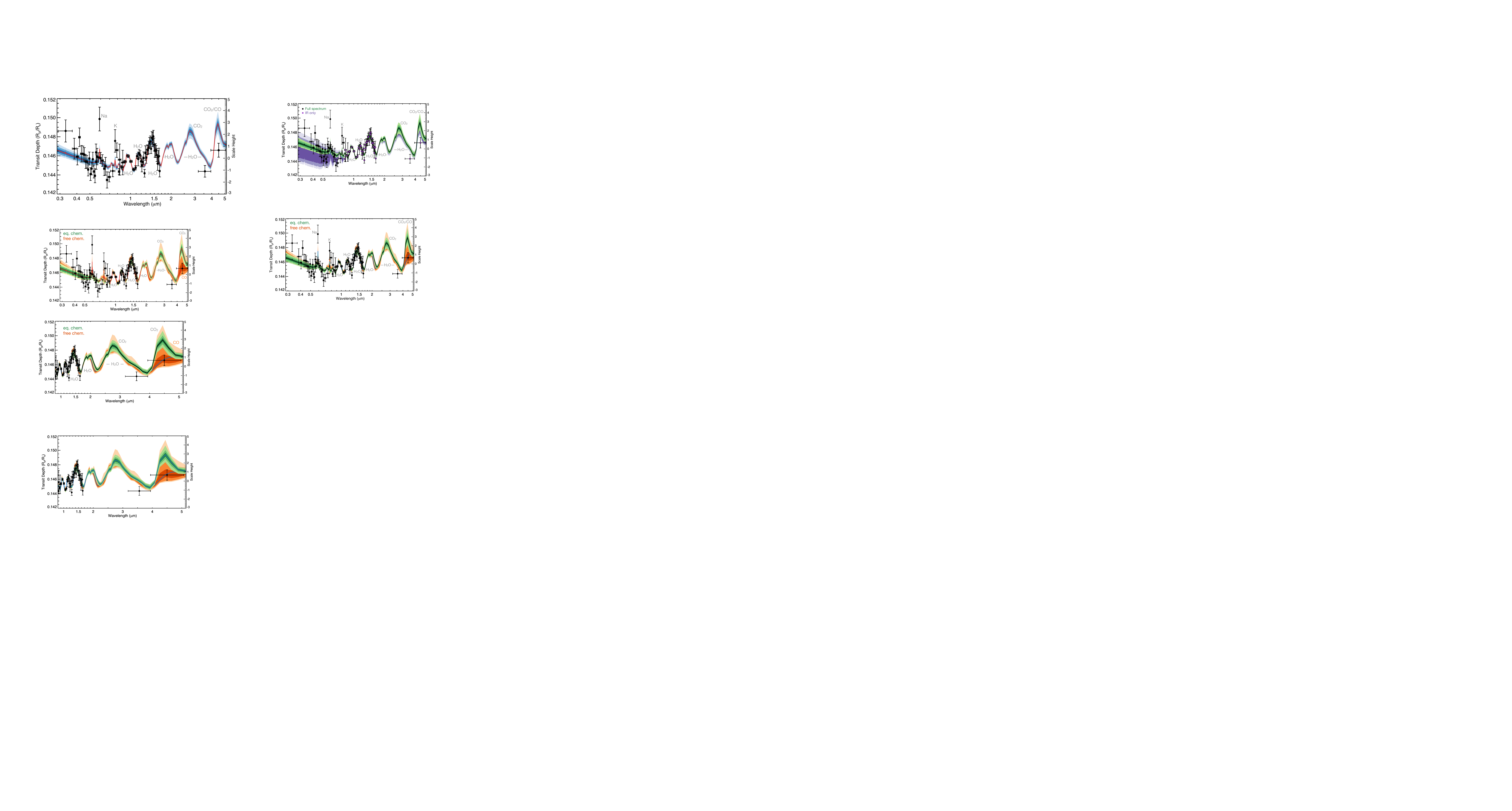}
\caption{The transmission spectrum of WASP-39b (black points). We show the result from both ATMO retrievals using equilibrium chemistry (green) and free chemistry (orange) with the 1, 2, and 3$\sigma$ bounds as in Fig.\ref{fig:W39_ARCeq_spec}.} 
\label{fig:W39_eq_v_free}
\end{figure}

The only current evidence of carbon-based species in the transmission spectrum of WASP-39b is in the 4.5\,$\mu$m \textit{Spitzer} bandpass, which covers potential absorption by CO and/or CO$_2$. In Fig.\,\ref{fig:W39_eq_v_free} we show the difference in retrieved models from the equilibrium and free-chemistry cases. From this we can see that the equilibrium chemistry model struggles to fit the data, as the fit is likely dominated by the high precision transmission spectrum below 1.7\,$\mu$m. In the free-chemistry model, because all the molecules are individually fit, the CO/CO$_2$ can be balanced such that each molecules is fit by the model. However, this results in much larger uncertainties on the data as wider distributions are invoked to account for low resolution data. The CO feature also extends slightly beyond the photometric wavelength range covered by the \textit{Spitzer} point, which also extends the uncertainty of the model. 
Due to the high metallicity of the atmosphere, the 4.5\,$\mu$m feature is likely dominated by CO$_2$ absorption (\citealt{moses2013}). Future observations with JWST will be able to better distinguish between carbon-based species in the atmosphere and it is then likely the free-chemistry model will provide a more informative retrieval. However, with the current data and based on the \corr{$\chi^2_\nu$ and} BIC, the equilibrium model has the greatest statistical significance; as such we use this model and results to further interpret the atmosphere of WASP-39b.

\begin{figure*}
\centering 
	\includegraphics[width=0.98\textwidth]{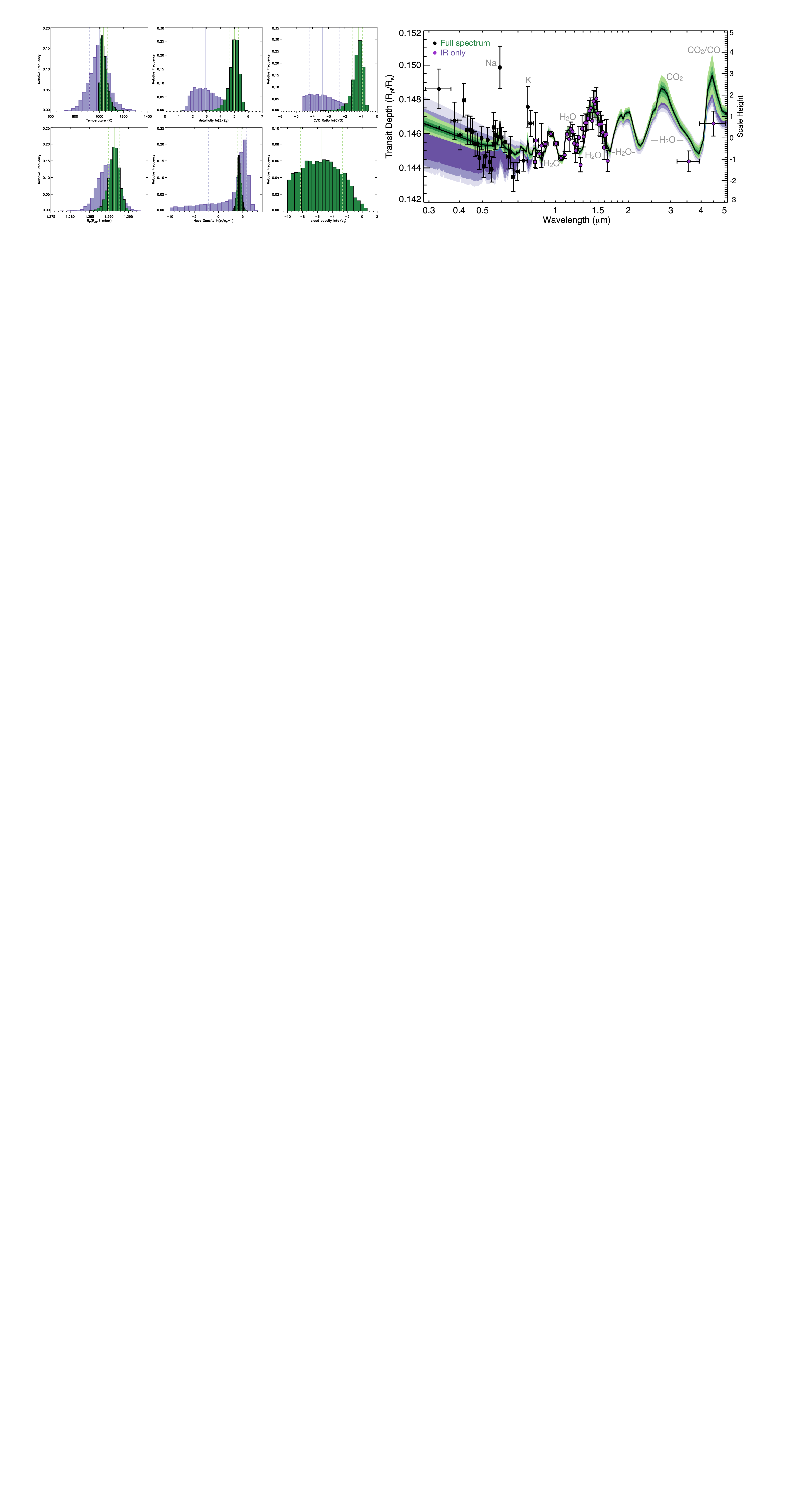}
\caption{Left: histograms showing the relative frequency distributions for each parameter (Temperature, [M/H], C/O, R$_p$/R$_*$, Haze, and Cloud) in the equilibrium retrieval for the full dataset (green) and just the IR portion of the dataset (purple). Right: the complete transmission spectrum of WASP-39b (black points), where the IR data is indicated by purple points. The two models show equilibrium models fit to the full transmission spectrum (green), and just the IR points (purple), with the 1, 2, and 3$\sigma$ bounds. These clearly shows the uncertainty associated with limited wavelength coverage and the importance of the optical. }
\label{fig:W39_optical_v_IR}
\end{figure*}

\subsection{The importance of the optical data in constraining molecular abundances}\label{opvir}
We next explored the impact different wavelength regions have on the overall metallicity constraint by performing an identical equilibrium retrieval, but excluding the optical HST and VLT data below 0.8\,$\mu$m. Optical transmission spectra have been theoretically shown to be important when measuring the volume mixing ratios of species (\citealt{Benneke2012,Griffith2014,line2016,Heng2017}), as an infrared-only transmission spectra provides only the relative abundances between molecules. However, few demonstrations are available using optical spectral data.  While similar data are available for other exoplanets (\citealt{Sing2016}), this test is particularly enlightening with WASP-39b, as it has continuous wavelength coverage from 0.3 to 1.7 $\mu$m thanks to the WFC3/G102 data. This complete coverage includes pressure-broadened alkali lines (\citealt{fischer2016}), multiple well-resolved water features, and overall higher quality data (precision and resolution) than many other exoplanetary spectra.

We use equilibrium chemistry models to compare the retrieved results for the full and partial WASP-39b data, as it provided the best overall fit to the data as measured by the BIC.  The results can be seen in Fig. \ref{fig:W39_optical_v_IR}.  Despite adopting chemical equilibrium and including the Spitzer IRAC data in the retrieval, the infrared-only data was unable to constrain the lower end of the C/O ratio, and a very strong degeneracy was observed between the C/O and [M/H].  Without the optical data, the 3-$\sigma$ range of the transmission spectra at 0.3$\mu$m encompasses 3.4 pressure scale heights (Fig. \ref{fig:W39_optical_v_IR} purple range) and contains a wide range of models, including ones with a strong haze, completely cloud- or haze-free models, and models with both near solar and highly super-solar metallicity (see Fig. \ref{fig:W39_optical_v_IR}).  The posterior of the infrared-only retrieval shows the [M/H] is measured to 0.64 dex, however, a lower prior value of 0.01 was enforced on the C/O ratio which, in turn, also limits the lower range of [M/H].  By including the optical HST and VLT data into the retrieval, pressure information from the alkali lines and constraints on the near-UV scattering slope help to limit the parameter space and [M/H], which in turn helps constrain the C/O ratio as well.  Even though the near-UV transmission spectral features can not uniquely be pinned to Rayleigh scattering of the bulk H$_2$ gas, as a scattering aerosol contribution is also present and included in the retrieval model, the optical data does exclude models with either very high or very low haze components, which helps to constrain the fitted parameters.  With the optical data, the resulting [M/H] is measured to 0.14 dex, which is a 0.5 dex improvement over excluding the optical data.  The abundance constraints of water (as determined by the [M/H] and C/O ratio see \S\ref{arc_retrieval}) are also significantly improved from precisions of $\pm$86$\times$ to $\pm$46$\times$ solar.

From this test, we demonstrate that the optical spectra with the data quality as provided by HST and VLT can provide an important contribution in constraining the abundances of molecules identified from infrared transmission spectra.  In the absence of optical data, breaking the [M/H] - C/O degeneracy will likely require complete near-infrared spectral coverage to identify all the major molecular components (CO, CO$_2$, CH$_4$, H$_2$O) such that the C/O ratio can be directly measured and in turn [M/H] constrained (\citealt{Greene2016}).  For JWST, coverage between 0.6 and 5 $\mu$m for targets brighter than J$\sim$10 will require at least two transit observations (e.g. NIRISS/SOSS and NIRSpec/G395H), while fainter targets can cover the range at low resolution in one transit with the NIRSpec prism.  Given that significant O and C could be locked up in other species such as condensates (\citealt{Greene2016}), the inclusion of optical data may also prove useful to identify potential biases when estimating the C/O ratio solely from the major molecular components.  Thus, significant leverage can be gained by combining JWST spectra with that of HST or other ground-based facilities, though care must always be taken when interpreting non-simultaneously gathered data, especially if the planet orbits an active star.   

\subsection{The impact of limb differences}\label{limb_differences}
Overall the retrieved temperature matches well with the 3D GCM model, where increases in the metallicity would shift the P-T profiles hotter (Fig. \ref{fig:TP}). At higher temperatures it might then be expected that the difference in temperature at the two limbs will also increase (\citealt{kataria2016}). Differences in limb temperatures may result in different conditions at the east and west limbs of the annulus such that cloud condensates can form at one and not the other (see Fig.\ref{fig:TP}). 

The presence of clouds on one limb of the planet but not the other can potentially mimic high metallicity signatures in the transmission spectrum (\citealt{line2016}). Using the simplistic toy model presented in \citet{line2016}, which uses a linear combination of two forward models to approximate limb differences, we test the scenario that the two limbs are different with a series of models on the full transmission spectrum of WASP-39b. We use a range of 1D isothermal models from the Goyal grid to represent different atmospheric scenarios on the planetary limbs. Using Fig. \ref{fig:TP} as a guide we select two models separated by 200\,K, where the cooler model (741\,K) represents the fully cloudy limb (i.e. a uniform optically thick cloud across all wavelengths), and the hotter model (966\,K) represents the clear, cloud-free limb. For each model set we keep the C/O, [M/H], and haze parameters the same. We test six sets with varied [M/H] values of 0.0, 1.7, and 2.0 dex (1$\times$, 50$\times$, and 100$\times$ solar). For each of the three metallicities we then test two different haze values, 10$\times$ and 150$\times$ (Table \ref{table:partly_cloudy}). We fit each separate model to the WASP-39b transmission spectrum by allowing it to move in altitude only, with all other parameters considered fixed. \corr{In effect, this only introduces one free parameter to the fit, however, it should be noted that the models themselves while fixed are based on a number of variables which when modeled more comprehensively to determine limb-differences may impact the statistics of fit to the data.} The $\Delta$BIC values presented in Table \ref{table:partly_cloudy} represent the difference in BIC between the best fit model and all other models in the table.
From this test we find that the higher metallicity models statistically fit the data better than low metallicity cases, even when 50/50 clear/cloudy models are considered. However, it should be noted that this is an oversimplified model and the parameter space being explored. Future observations of WASP-39b, which will likely have complete wavelength coverage, and much higher precision and resolution data, will likely require sophisticated 3D modeling to accurately infer anything further from the data when considering partly cloudy scenarios.

\begin{table}
\footnotesize{
\centering
\caption[\quad Partly Cloudy models]{Test cases for the impact of limb differences on the observed transmission spectrum. In each pair the two models contribute 50\% each to the combined model. Each model and the combined model are fit to the transmission spectrum and the $\chi^2$, BIC, and $\Delta$BIC (compared to the lowest BIC model) are calculated (see \S\ref{limb_differences})}
\begin{tabular}{cccccccc}
\hline
\hline
Group & T$_{eq}$  & [M/H] & Haze & Cloud & $\chi^{2}$ & BIC & $\Delta$BIC\\
 ~ & (K) & ~ & ~ & ~ & ~ & ~ & \\
\hline
\multirow{2}{*}{a)} & 741 & 0.0 &  10.0 & 1.0 & 196 & 200 & 50\\
& 966 & 0.0 &  10.0 & 0.0 & 327 & 331 & 181\\
50/50 &  &  &  &  & 179 & 187 & 37 \\
\hline
\multirow{2}{*}{b)} & 741 & 0.0 & 150.0 & 1.0 & 360 & 364 & 214 \\
& 966 & 0.0 &  150.0 & 0.0 & 757 & 762 & 612\\
50/50 &  &  &  &  & 514 & 522 & 372\\
\hline
\multirow{2}{*}{c)} & 741 & 1.7 &  10.0 & 1.0 & 149 & 153 & 3\\
& 966 & 1.7 &  10.0 & 0.0 & 326 & 330& 180\\
50/50 &  &  &  &  & 166 & 174 & 24 \\
\hline
\multirow{2}{*}{d)} & 741 & 1.7 &  150.0 & 1.0 & 180 & 184 & 34\\
& 966 & 1.7 &  150.0 & 0.0 & 224 & 228 & 78\\
50/50 &  &  &  &  & 174 & 182 & 32\\
\hline
\multirow{2}{*}{e)} & 741 & 2.0 &  10.0 & 1.0 & 157 & 161 & 11\\
& 966 & 2.0 &  10.0 & 0.0 & 224 & 228 & 78\\
50/50 &  &  &  &  & 142 & 150 & 0 \\
\hline
\multirow{2}{*}{f)} & 741 & 2.0 &  150.0 & 1.0 & 175 & 178 & 28\\
& 966 & 2.0 &  150.0 & 0.0 & 158 & 162 & 12 \\
50/50 &  &  &  &  & 146 & 154 & 2\\
\hline
\end{tabular}
\label{table:partly_cloudy}
}
\end{table}

\begin{figure*}
\centering 
	\includegraphics[width=0.96\textwidth]{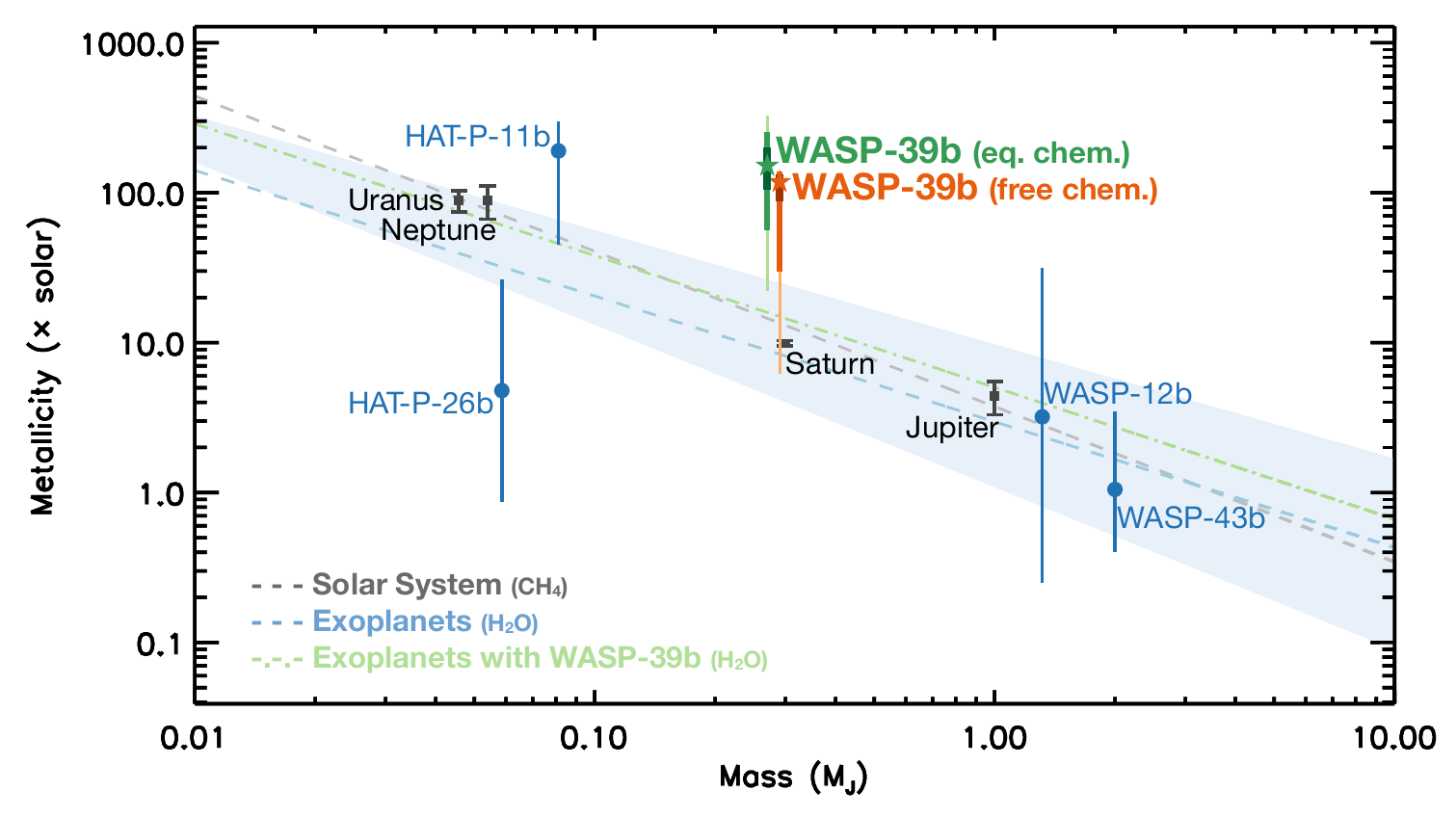}
\caption{Mass-metallicity relation for the solar system and exoplanets. We show the measured metallicities of the four giant planets in our solar system (black squares) fit for the methane abundance (grey dashed line), and four previously published exoplanets (blue circles) fit for the water abundance (blue dashed line)\corr{; all plotted metallicities show the 68\% confidence interval}. The shaded region represents the 1$\sigma$ diversity from all eight measured [M/H] and uncertainties. We show the retrieved metallicity of WASP-39b from the equilibrium chemistry fit (green left) and the free-chemistry fit (orange right) based on the water abundance of the atmosphere, with the \corr{68\%, 95\%, and , 99.7\% confidence intervals} (dark, medium, and light errorbars respectively). \corr{We also compute the fit to the exoplanet data including the WASP-39b equilibrium chemistry model using the 68\% confidence interval measurement (green dash-dot line).} WASP-39b metallicity results from each model are offset in mass by 0.01\,M$_J$ for clarity.}
\label{fig:W39_mass_metallicity}
\end{figure*}

These simple toy model results, along with the abundance measurements from equilibrium and free chemistry retrievals, further suggest that WASP-39b has a high metallicity atmosphere. We detail the implications of this in the following section.

\subsection{The atmospheric metallicity of WASP-39b}\label{metallicity}
There are two scenarios generally considered for the formation of giant planets prior to the assumed migration of close-in giant exoplanets: gravitational instability and core accretion. Planetary atmospheres will exhibit different atmospheric properties under formation via these two formation pathways. Gravitational instability theory suggests that planets will have the same atmospheric metallicity as the central star, while with core accretion theory, lower mass planets will have higher atmospheric metallicity (e.g. \citealt{mordasini2012,fortney2013}). Studies of our solar system giant planets fit with the predicted trend of core accretion when the atmospheric methane abundance is used as a proxy for overall metallicity (c.f \citealt{kreidberg2014b}). More recently, atmospheric water absorption features in exoplanet atmospheres have been used with retrieval modeling to constrain the overall atmospheric metallicity using oxygen as a proxy for the heavy element abundance (\citealt{kreidberg2014b,fraine2014,kreidberg2015,Wakeford2017science}). The first measurement of this type was conducted on WASP-43b (\citealt{kreidberg2014b}) which directly fit with the trend established by the solar system supporting core accretion theory prior to inward migration. However, in a more recent study of the Neptune-mass exoplanet HAT-P-26b (\citealt{Wakeford2017science}), we showed a deviation from this trend in the low mass regime, hinting at diversity in formation location and/or time. This is consistent with the envelope accretion models by \citet{leechiang2016}, which argue that most hot Neptunes accrete their envelopes in-situ shortly before the disk dissipates resulting in lower heavy element contamination in the atmosphere.

To better approximate the correlation in mass-metallicity space, we separately calculate linear fits to the methane and water abundance measurements for the four solar system giant planets, and the four published exoplanet measurements respectively.   
To qualitatively assess the significance of fit to the data we use the \corr{$\rho^2$ statistic (\citealt{mcfadden1974}), defining it here} as,
\begin{equation}
\rho^2 = 1 - \frac{\sum{\frac{(x-line)^2}{\sigma^2_x}}}{\sum{\frac{(x-mean)^2}{\sigma^2_x}}} = 1 - \frac{\chi^2_{line}}{\chi^2_{mean}} 
\end{equation}
where, $x$ is the data, $line$ is the linear fit to the data, $mean$ is the mean of the data, and $\sigma_x$ is the uncertainty on the data \corr{assuming a Gaussian distribution with symmetric uncertainties in log-metallicity space}.
The \corr{$\rho^2$} statistic \corr{evaluates the improvement that the more complex model has to the fit, compared with a more simplistic model. In this case, it balances the likelihoods of the data being drawn from a model where there is a correlation between mass and metallicity and the data being drawn from a model with no correlation, fixed at the average metallicities of the data}.
From this statistic we find that $\sim$93\% of the scatter observed in the solar system mass-metallicity relation can be explained by a linear model, even when the uncertainties are taken into account. For the previously published exoplanet data (Fig.\ref{fig:W39_mass_metallicity}, blue circles) we find that \corr{60\%} of the variance can be explained by a linear fit to the data.

Using the water abundance as a proxy for overall atmospheric metallicity, we constrain the atmospheric metallicity of WASP-39b to be 151$^{+48}_{-46}$\,$\times$solar, at \corr{the 68\% confidence interval,} from a retrieval using equilibrium chemistry. We show the metallicty of WASP-39b relative to other giant planets in Fig. \ref{fig:W39_mass_metallicity}. We also include the \corr{95\% and 99.7\%} bounds of the retrieved metallicity from the equilibrium and free-chemistry fits to better demonstrate the similarities and bounds of each retrieval.
When the new WASP-39b results are incorporated into the exoplanet fit, the \corr{$\rho^2$ statistic drops significantly to just 24\% statistical association with a linear fit.} This does not rule out a linear fit to the exoplanet data, \corr{and indeed the exoplanet linear fits both have log Bayes factors on the order of 5, suggesting a tentative positive relationship (\citealt{Kass1995}). It} merely suggests that more high precision data are required to determine a trend in mass-metallicity space. 
As shown by the \corr{$\rho^2$ statistic}, with each new exoplanet metallicity measurement, it should be expected that the linear fit of the mass-metallicity relation will evolve in both variance and constraint. These future observations may also show that systems with multiple giant planets provide a more telling comparison to the solar system, as multi-planet systems may have entirely different metallicities than single-planet systems.
\begin{table}
\centering
\caption{Statistical significance of the mass-metallicity plot \corr{following Eqn. 6.}} 
\begin{tabular}{lcc}
\hline
\hline
Data & \corr{$\rho^2$} \\
\hline
Solar System & 0.93 \\
Exoplanets (no WASP-39b) & \corr{0.60} \\
Exoplanets (with WASP-39b) & \corr{0.24} \\
\hline
\end{tabular}
\vspace{-1pt}
\label{table:stats}
\end{table}
Simulations presented in \citet{thorngren2016} show that the scatter in the heavy element fraction relative to the mass of gas giant planets is expected to be large. This suggests that the metallicity measurement for WASP-39b retrieved here is not entirely unexpected, although possibly at the extreme top edge of the scatter. Following the core accretion theory, this suggests that WASP-39b formed in a region rich with heavy element planetesimals, likely in the form of ices, which were accreted by the planet during formation. The potential presence of ices implies formation beyond the ice lines of the planet-forming disk, more akin to Neptune and Uranus's orbital distance than that of the similar mass planet Saturn in our own solar system.  

\section{Conclusion}
We present the complete transmission spectrum of the Saturn-mass exoplanet WASP-39b by introducing new measurements between 0.8--1.7\,$\mu$m using HST WFC3. We measure distinct water absorption over three bands with a maximum base to peak amplitude of 2.4 planetary scale heights (H) and an average amplitude of 1.7\,H. Using the ATMO Retrieval Code (ARC) we constrain the temperature to 1030$^{+30}_{-20}$\,K, and  the atmospheric metallicity at 151$^{+48}_{-46}\times$\,solar, based predominantly on the water abundance. At a 1$\sigma$ confidence this still represents significant diversity from the current mass-metallicity trends based on either atmospheric methane or water abundance for giant planets. This suggests that WASP-39b formed beyond the snow line in the planet-forming disk of the host star, where it likely accumulated metal-rich ices and planetesimals prior to later inward migrations to its current orbital position. 
However, overall more precise exoplanet abundances are be needed before definitive conclusions can be made with regards to the exoplanet mass-metallicity relation and planetary formation pathways.

WASP-39b is an ideal target for follow-up studies with the James Webb Space Telescope (JWST) to precisely measure the atmospheric carbon species and abundance already hinted at in these early investigations. We predict that due to the high metallicity of WASP-39b's atmosphere, CO$_2$ will be the dominant carbon species. This can be measured at 4.5$\mu$m with JWST NIRSpec G395H, allowing further constraint to be placed on the C/O and atmospheric metallicity.

%
\section{Acknowledgements}
The authors thank K.B.\,Stevenson for useful discussions on the data analysis. This work is based on observations made with the NASA/ESA Hubble Space Telescope that were obtained at the Space Telescope Science Institute, which is operated by the Association of Universities for Research in Astronomy, Inc. These observations are associated with programs GO-14169 (PI. H.R.\,Wakeford) and GO-14260 (PI. D.\,Deming).
D.K.\,Sing, H.R.\,Wakeford, T.\,Evans, B.\,Drummond, N.\,Nikolov, acknowledge funding from the European Research Council (ERC) under the European Union’s Seventh Framework Programme (FP7/2007-2013) / ERC grant agreement no. 336792. J.\,Goyal acknowledges support from Leverhulme Trust. A.L.\,Carter acknowledges support from the STFC. 
H.R.\,Wakeford also acknowledges support from the Giacconi Fellowship at the Space Telescope Science Institute, which is operated by the Association of Universities for Research in Astronomy, Inc. 
This research has made use of NASAs Astrophysics Data System, and components of the IDL astronomy library, and the Python modules SciPy, NumPy, and Matplotlib.
Many thanks go to the crew of STS-125 HST servicing mission 4, for fixing HST and for installing WFC3 over a period of 5 EVAs that took a total of 36 hours 56 minutes, almost matching the total exposure time taken by these observations. Also, thank you to Mac Time machine without which this project would not have been possible, due to multiple moves and hard-drive failures.\\


{\it Facilities:} \facility{HST (WFC3)}. \\
%
\bibliographystyle{mn2e}
\bibliography{W39_references.bib}

\end{document}